%% file: main.tex
\documentclass[twocolumn]{aastex631}
\usepackage{amsmath} 
\usepackage{soul} 
\usepackage{multirow} 
\usepackage{bm} 
\usepackage{amsmath} 
\usepackage{xspace} 
\usepackage{gensymb} 
\usepackage{float}
\usepackage{customcommands} 
\usepackage{newunicodechar,graphicx}
\usepackage{multirow, bigdelim, makecell, booktabs}
\PassOptionsToPackage{hyphens}{url}\usepackage{hyperref}

\DeclareRobustCommand{\okina}{%
  \raisebox{\dimexpr\fontcharht\font`A-\height}{%
    \scalebox{0.8}{`}%
  }%
}
\newunicodechar{ʻ}{\okina}

\begin{document}
\title{The impact of observing cadence and undetected companions on the accuracy of planet mass measurements from radial velocity monitoring}
\correspondingauthor{Joseph M. Akana Murphy}
\email{joseph.murphy@ucsc.edu}
\input{author_info}
\begin{abstract}
We conduct experiments on both real and synthetic radial velocity (RV) data to quantify the impact that observing cadence, the number of RV observations, and undetected companions all have on the accuracy of small planet mass measurements. We run resampling experiments on four systems with small transiting planets and substantial public data from HIRES in order to explore how degrading observing cadence and the number of RVs affects the planets' mass measurement relative to a baseline value. From these experiments, we recommend that observers obtain 2--3 RVs per orbit of the inner-most planet and acquire at minimum 40 RVs. Following these guidelines, we then conduct simulations using synthetic RVs to explore the impact of undetected companions and untreated red noise on the masses of planets with known orbits. While undetected companions generally do not bias the masses of known planets, in some cases, when coupled with an inadequate observing baseline, they can cause the mass of an inner transiting planet to be systematically overestimated on average.
\end{abstract}

\keywords{Exoplanets (498), radial velocity (1332)}
\section{Introduction} \label{sec:intro}
Time series Doppler monitoring of transiting exoplanet host stars is the workhorse technique for populating the exoplanet mass-radius diagram. For planets discovered in 2018 or later \citep[i.e., the year that NASA launched the Transiting Exoplanet Survey Satellite or \tess;][]{ricker14}, 185 of the 194 precisely characterized\footnote{Planets with better than 33\% and 15\% fractional measurement precision in mass and radius, respectively \citep{nea}.} planets smaller than 4 \rearth have masses that were measured using the radial velocity (RV) method---the other 9 masses were measured with transit timing variations \citep[TTVs; e.g.,][]{hamann19}. The precise masses of super-Earths (1 \rearth $\lesssim$ \rplanet $\lesssim$ 1.8 \rearth) and sub-Neptunes (1.8 \rearth $\lesssim$ \rplanet $\lesssim$ 4 \rearth) are particularly valuable, as they set the stage for detailed atmospheric follow-up observations \citep{batalha19}, the results of which may break degeneracies \citep{rogersSeager10} that currently exist between theoretical models of small planet composition \citep[e.g.,][]{valencia07, adams08, aguichine21, rogers23}.

As the community turns toward studying the relative demographics of small planets in the mass-radius plane \citep[e.g.,][]{otegi20a, luque22, polanski24}, mass measurement accuracy will confuse our understanding of the observed mass-radius distribution. Kepler-10 c is an example of a short-period, small planet ($P = 45.3$ days, \rplanet $= 2.3$ \rearth) whose mass measurement changed drastically (by more than $3\sigma$) upon further observation \citep{dumusque14, weiss16, rajpaul17}. Another particularly concerning instance of mass measurement inaccuracy is the case of \sysI b ($P = 17$ days, \rplanet $= 2.63 \pm 0.11$ \rearth), which was previously though to be one of the most massive sub-Neptunes known \citep[\mplanet $=$ \mpIbLXIX \mearth;][]{luque19}. However, additional RV follow-up recently revealed that the planet's mass is, in fact, nearly a third of its original value: \mplanet $=$ \mpIb \mearth \citep[\mplanet $<$ \mpIbupperlim \mearth at 98\% confidence;][]{hd119130murphy24arxiv}.

Even relatively small changes (by a few \mearth) in the masses of super-Earths and sub-Neptunes can have important implications for their bulk composition, and, therefore, scenarios of their formation and evolution \citep{luque22, cadieux24}. Another concern is the potential misinterpretation of transmission spectra or discrepancies between the expected and observed signal-to-noise ratios (\snr) of molecular features due to an incorrect planet surface gravity estimate. Inaccurate masses may also bias mass-radius relations for small planets \citep{burt18}, and propagate the issue to future Doppler observing proposals that make use of their results for exposure time estimation and/or target selection \citep{burt24}. 

The topic of inaccurate RV mass measurements and optimal RV observing strategies has not gone without comment. \cite{cochran96} presented one of the earliest studies of RV detection efficiency with respect to the number of observations obtained from a uniform sampling of orbital phase. More recently, much attention has been paid to the influence of orbital eccentricity on measured planet RV semi-amplitudes \citep{shen08, otoole09, rodigas09}. Optimal observing strategies have also been proposed using synthetic RV observations \citep{burt18, gupta24}. In this study, we explore the impacts that observing cadence, the number of RV measurements, and the presence of undetected companions have on the accuracy of planet masses using both real and synthetic RV data. 

The paper is organized as follows. In \S\ref{sec:data}, we describe the public \keckhires data used in this work. In \S\ref{sec:real_data}, we explore how observing cadence and the number of observations affect the accuracy of planet mass measurements for four illustrative systems. In \S\ref{sec:synth_data}, we use simulations of synthetic data to quantify the impact of undetected signals on the mass measurement of a planet with a known orbit. We summarize our results in \S\ref{sec:conclusion}.

\section{Data}\label{sec:data}
Here we briefly describe the public data from \keckhires that we use in our mass measurement accuracy experiments (\S\ref{sec:real_data}). The High Resolution Echelle Spectrometer \citep[HIRES;][]{vogt94} is mounted on the 10\,m Keck I telescope at the W. M. Keck Observatory on Maunakea, Hawai\okina i. HIRES RVs are computed using the iodine method, in which observations are taken with a warm (50\degree\ C) cell of molecular iodine at the entrance slit \citep{butler96}. The superposition of the iodine absorption lines on the stellar spectrum provides both a fiducial wavelength solution and a precise, observation-specific characterization of the instrument's point spread function (PSF). 

We make use of four archival \keckhires time series of transiting planet hosts observed by the \tess-Keck Survey \citep[TKS;][]{chontos22, polanski24}. The systems are \sysII \citep{murphy23}, \sysIII \citep[aka TOI-266;][]{murphy23}, \sysIV \citep{dai21}, and \sysV \citep{macdougall22}. In addition to their original confirmation papers, these systems also have mass measurements from a homogeneous analysis of all TKS data \citep{polanski24}. In our mass measurement accuracy analysis (\S\ref{sec:real_data}), we use the RV data from \cite{polanski24}, since they represent the complete set of TKS observations. For \sysII and \sysIII, the RV data sets used in \cite{polanski24} and \cite{murphy23} are the same. For \sysIV, \cite{polanski24} includes 22 additional \keckhires RVs compared to \cite{dai21}. For \sysV, \cite{polanski24} includes 14 additional \keckhires RVs compared to \cite{macdougall22}. All of the RV data reduction followed the methods of \cite{howard10}.

\section{Mass measurement accuracy as a function of sampling cadence and the number of observations} \label{sec:real_data}
To gain a qualitative understanding of how sampling cadence, the number of observations, and model misspecification all affect the accuracy of RV mass measurements of small transiting planets, we conducted case studies of four \tess systems observed with \keckhires by the TKS. We chose to use archival TKS data because it was collected and reduced uniformly, and we selected systems with only \keckhires RVs to freeze out systematic differences between instruments. We resampled the \keckhires data to a grid in the number of RVs and minimum observing cadence (i.e., a lower limit on the time separation between consecutive observations). We then fit these resampled data sets and compared each planet's resulting best-fit $K$ to a baseline value obtained by fitting the complete \keckhires time series. This resampling procedure was conducted multiple times using different model assumptions to assess how model misspecification might impact the results (e.g., in some cases we purposely fit only a single Keplerian to the RVs of a multi-transiting system). The experiments are described in further detail in \S\ref{sub:real_data_methods}.

\begin{figure}
    \centering
    \includegraphics[width=\columnwidth]{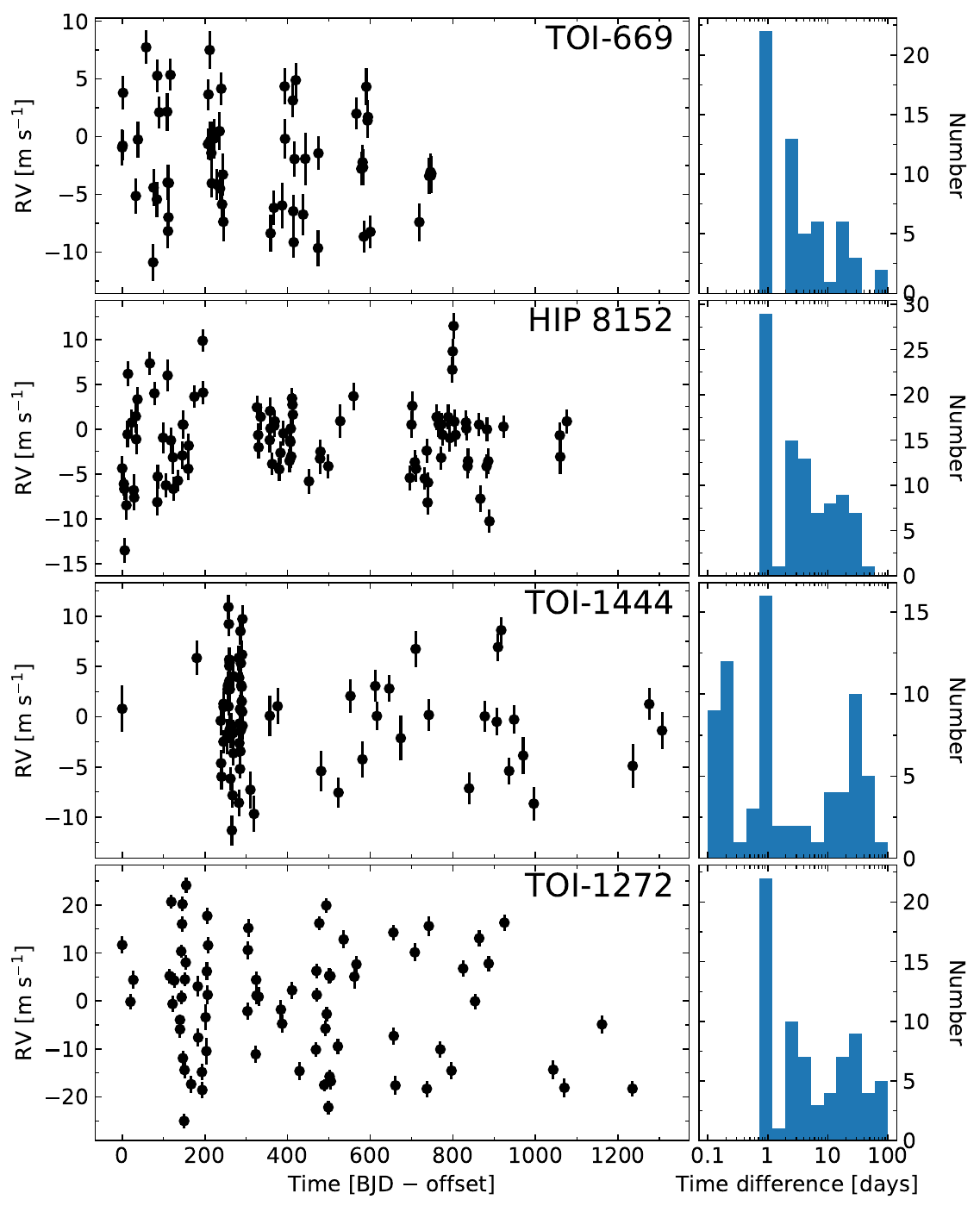}
    \caption{For each row: \emph{Left:} The \keckhires RV time series from \cite{polanski24}. Data are shown with an arbitrary offset such that the first observation is plotted at Time $= 0$. RVs are binned in units of 0.1 days. \emph{Right:} Histogram of the time differences between consecutive RVs. Note the intra-night cadence for \sysIV, which hosts an ultra-short period planet.}
    \label{fig:rv_timeseries}
\end{figure}

The four systems---\sysII, \sysIII, \sysIV, and \sysV---were chosen because they are all of similar spectral type (\teff $=$ 5000--5600 K), are all host to at least one small (\rplanet $< 4.5$ \rearth) transiting planet, and all have large (\nrv $> 60$), public \keckhires RV data sets which were acquired by a single survey following the same observing and data reduction procedures. Figure \ref{fig:rv_timeseries} shows the RV time series for each system, as well as the distribution of the time differences between consecutive RVs. \sysII, \sysIII, and \sysIV were also chosen because they are all chromospherically inactive (\logrhk $< -5.00$). We included \sysV for comparison because it is slightly active, with \logrhk $=$ \logrhkV. 

We note that although our analysis in this section is independent of host star properties,\footnote{The only instance where our analysis depends on the property of a host star is the center of the Gaussian prior placed on the hyperparameter corresponding to the stellar rotation period for the Gaussian process component used in some of our models of the \sysV RVs. See Table \ref{tab:radvel_model_real_data}.} we summarize fundamental stellar properties for these systems in Table \ref{tab:case_studies_stellar_properties} for ease of reference.

\input{case_studies_stellar_properties.tex}

\subsection{Methodology} \label{sub:real_data_methods}
For a system with $N_\mathrm{tot}$ RV measurements, we resampled the RV time series to a grid in the number of RVs included in the fit (\nrv) and the minimum observing cadence (i.e., the minimum time separation allowed between consecutive observations; MOC). The grid is defined by \nrv $=$ [10, 11, 12, ..., $N_\mathrm{tot}$] and MOC $=$ [0, 1, 2, ..., 30] days, where the coordinate pair (\nrv $= N_\mathrm{tot}$, MOC $= 0$) represents the full data set. Note that a given data set cannot necessarily support all combinations of \nrv and MOC on the grid, so these cells are excluded from the resampling and fitting steps. In practice, to generate each resampled data set, we start with the earliest observation in the complete time series and add subsequent observations according to the specified MOC until we reach the desired \nrv.

These resampled data sets were each fit with an RV model using \radvel \citep{radvel}---the ``fitting'' model. The resulting $K$ of each planet as measured by the fitting model was compared to the corresponding value from a fit to the full data set using the system's ``baseline'' model. For each system, we first picked the fitting model to be the baseline model to isolate the effects of the data resampling on the recovered $K$ values. In subsequent experiments, we explored other configurations of the fitting model to assess the impacts of model (mis)specification. 

To choose the baseline model for each system, we computed and compared the small sample-corrected Akaike Information Criterion \citep[AICc;][]{akaike74, burnham02} for a variety of models. The AICc is defined as 
\begin{equation} \label{eqn:aicc}
    \mathrm{AICc} = \mathrm{AIC} + \frac{2 N_\mathrm{par} (N_\mathrm{par} + 1)}{N_\mathrm{obs} - N_\mathrm{par} - 1},
\end{equation}
and the Akaike Information Criterion \citep[AIC;][]{akaike74} is given as
\begin{equation} \label{eqn:aic}
    \mathrm{AIC} = 2 N_\mathrm{par} - 2 \ln \mathcal{\hat{L}}.
\end{equation}
$N_\mathrm{par}$ is the number of free model parameters, $N_\mathrm{obs}$ is the number of observations, and $\mathcal{\hat{L}}$ is the model's likelihood function maximized with respect to the model parameters. Throughout, we assume a log-likelihood of the form 
\begin{equation} \label{eqn:likelihood}
    \ln \mathcal{L} = -\frac{1}{2} \sum_{i = 1}^{N_\mathrm{obs}} \Big[\ln (2 \pi \sigma_{i}^2) + \frac{(x_i - \mu_{i})^2}{\sigma_{i}^2}\Big],
\end{equation}
which describes observations taken with a single instrument and where $x_i$ is the RV of the $i$-th observation taken at time $t_i$. We define
\begin{equation} \label{eqn:jitter}
    \sigma_{i}^2 = \sigma_{\mathrm{RV},\:i}^2 + \sigma_\mathrm{jit}^2,
\end{equation}
where $\sigma_{\mathrm{RV},\:i}$ is the measurement uncertainty on the $i$-th observation and $\sigma_\mathrm{jit}$ is the corresponding instrumental RV jitter parameter, which acts as an additional white noise term to compensate for otherwise untreated instrumental systematics. The model's predicted RV at time $t_i$ is given as
\begin{equation} \label{eqn:mean}
    \mu_{i} = \mathrm{Keplerian}(t_i,\: P,\:T_\mathrm{c},\:e,\:\omega_\mathrm{p},\:K) + \gamma
\end{equation}
where Keplerian(...) is the sum of the RV contributions from all of the Keplerian orbits considered by the model and $\gamma$ is the instrumental RV offset.

While there are no hard and fast rules that dictate model selection with the AIC (or AICc), \cite{burnham04} provide the following guidelines for comparing the AIC of a reference model, AIC$_0$, and of some other model, AIC$_i$: 
\begin{itemize}
    \item If $\mathrm{AIC}_i - \mathrm{AIC}_0 \equiv \Delta \mathrm{AIC}_i < 2$, the two models are nearly indistinguishable.
    \item If $2 < \Delta \mathrm{AIC}_i < 10$, the $i$th model is disfavored.
    \item If $\Delta \mathrm{AIC}_i > 10$, the $i$th model is essentially ruled out.
\end{itemize}

The AICc is essentially the same as the AIC, but with an additional penalty term for model complexity, since the AIC has a tendency to favor models that overfit when $N_\mathrm{obs}$ is small. \cite{burnham02} recommend using the AICc in place of the AIC when $N_\mathrm{obs} \lesssim 40 \times N_\mathrm{par}$, which is typically the case for models of RVs, where $N_\mathrm{par}$ is usually on the order of 10 and $N_\mathrm{obs} \lesssim 100$. 

In comparing different model configurations, we considered RV models that included orbital eccentricity, included background trends (long-term linear and/or quadratic trends), excluded any outer nontransiting planet candidates (for \sysIV and \sysV), and models resulting from different combinations of these choices. In each case, we found that a model that treats the orbit of each planet as being circular (save for \sysV b), that does not include linear or quadratic trends, and, for \sysIV and \sysV, that includes the nontransiting candidates, is either the AICc-preferred model itself or has $\Delta$AICc $< 5$ with the AICc-preferred model. These models were therefore chosen as the ``baseline'' models.

Table \ref{tab:radvel_model_real_data} describes the baseline model for each system. As expected, these models are the same as the adopted RV models from the original confirmation papers.\footnote{In the case of \sysIV, \cite{dai21} also include a linear RV trend in their adopted model, but they note that it is marginally disfavored by the Bayesian Information Criterion \citep[BIC;][]{schwarz78} and its posterior distribution is consistent with zero.} In the case of \sysV, we also explore a model that includes a Gaussian process \citep[GP; e.g.,][]{rasmussen06} component to account for correlated stellar activity.
\input{radvel_model_real_data.tex}

\subsection{\sysII: One transiting planet} \label{sec:toi669}
\sysII is an inactive G dwarf known to host a transiting sub-Neptune, \sysII b, on an orbit of $P = 3.95$ days with \rplanet $=$ \rpIIb \rearth and \mplanet $=$ \mpIIb \mearth \citep{murphy23}. The orbit of the planet is consistent with being circular according to a joint model of the \tess photometry and the \keckhires RVs, and the planet's RV semi-amplitude is $4.30 \pm 0.62$ \mps \citep{murphy23}. The \keckhires data comprises 62 RVs spanning a baseline of 748 days. 

\begin{figure}
    \centering
    \includegraphics[width=\columnwidth]{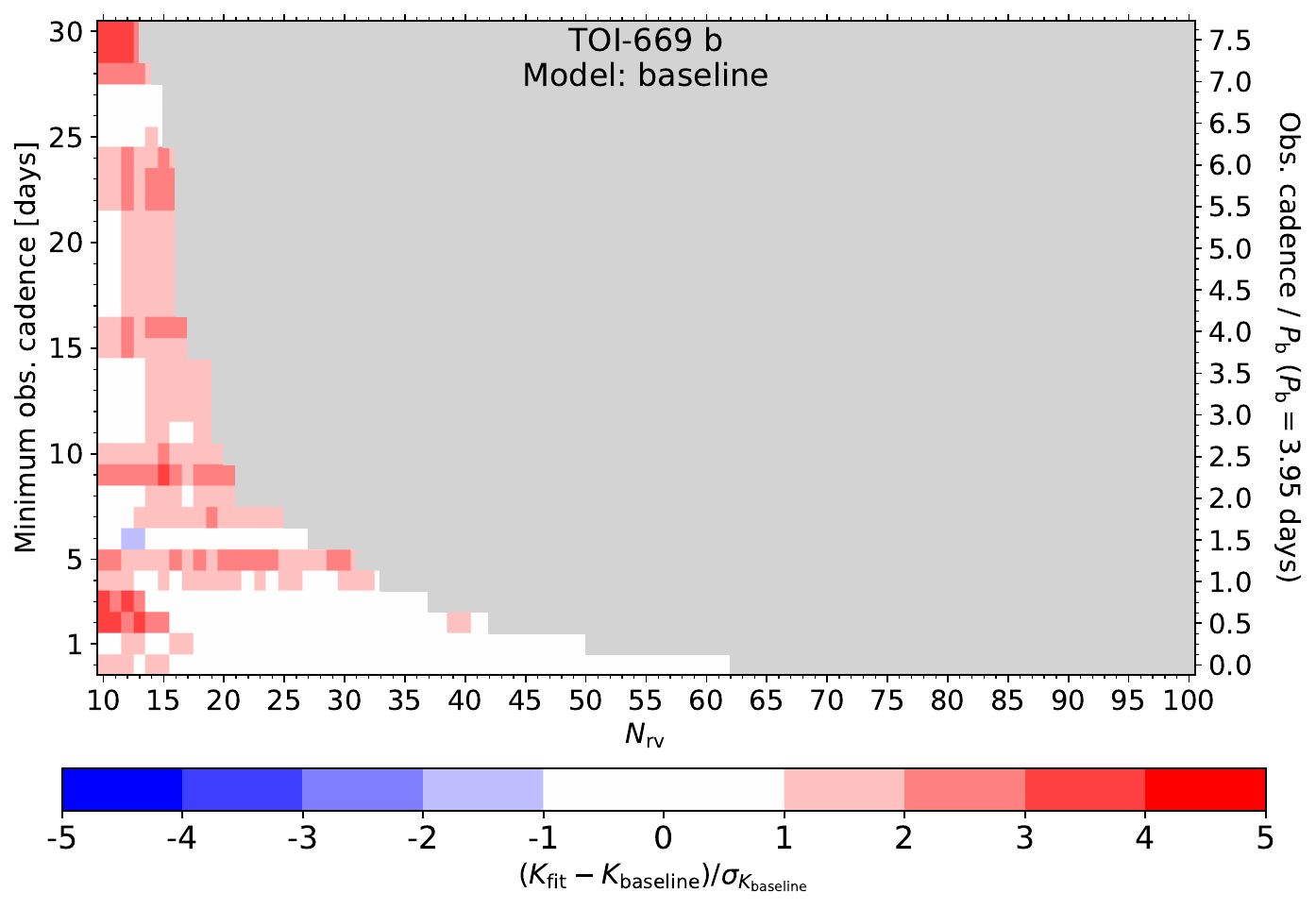}
    \caption{The results of the resampling experiment for \sysII b. Each cell represents a resampled realization of the full data set according to specified MOC (left y-axis) and \nrv (x-axis) values. The right y-axis is the same as the left, but it is now shown in units of the orbital period of the planet. The color of each cell represents the difference between the planet's measured semi-amplitude using that cell's resampled data set, $K_\mathrm{fit}$, and the measured semi-amplitude of the planet from the baseline model ($K_\mathrm{baseline}$), in units of the error on $K_\mathrm{baseline}$. Gray regions represent combinations of MOC and \nrv that are not supported by the data.}
    \label{fig:t000669_real_base}
\end{figure}

We find that the RV semi-amplitude measurement for \sysII b is generally consistent with the baseline value (i.e., within $2\sigma$) for MOC $\lesssim 5$ and $N_\mathrm{RV} \gtrsim 20$. Figure \ref{fig:t000669_real_base} shows the results of the resampling experiment. 

\subsection{\sysIII: Two transiting planets} \label{sec:hip8152}
\sysIII is an inactive G dwarf known to host two transiting planets in a near 2:1 mean-motion resonance (MMR). According to \cite{murphy23}, \sysIII b has $P = 10.75$ days, \rplanet $=$ \rpIIIb \rearth, and \mplanet $=$ \mpIIIb \mearth, and \sysIII c has $P = 19.61$ days, \rplanet $=$ \rpIIIc \rearth, and \mplanet $=$ \mpIIIc \mearth. Both orbits are consistent with being circular, and planet b and c have RV semi-amplitudes of $2.42 \pm 0.53$ \mps and $2.40 \pm 0.55$ \mps, respectively \citep{murphy23}. The \keckhires data comprises 94 RVs spanning a baseline of 1076 days. 

\begin{figure}
    \centering
    \includegraphics[width=\columnwidth]{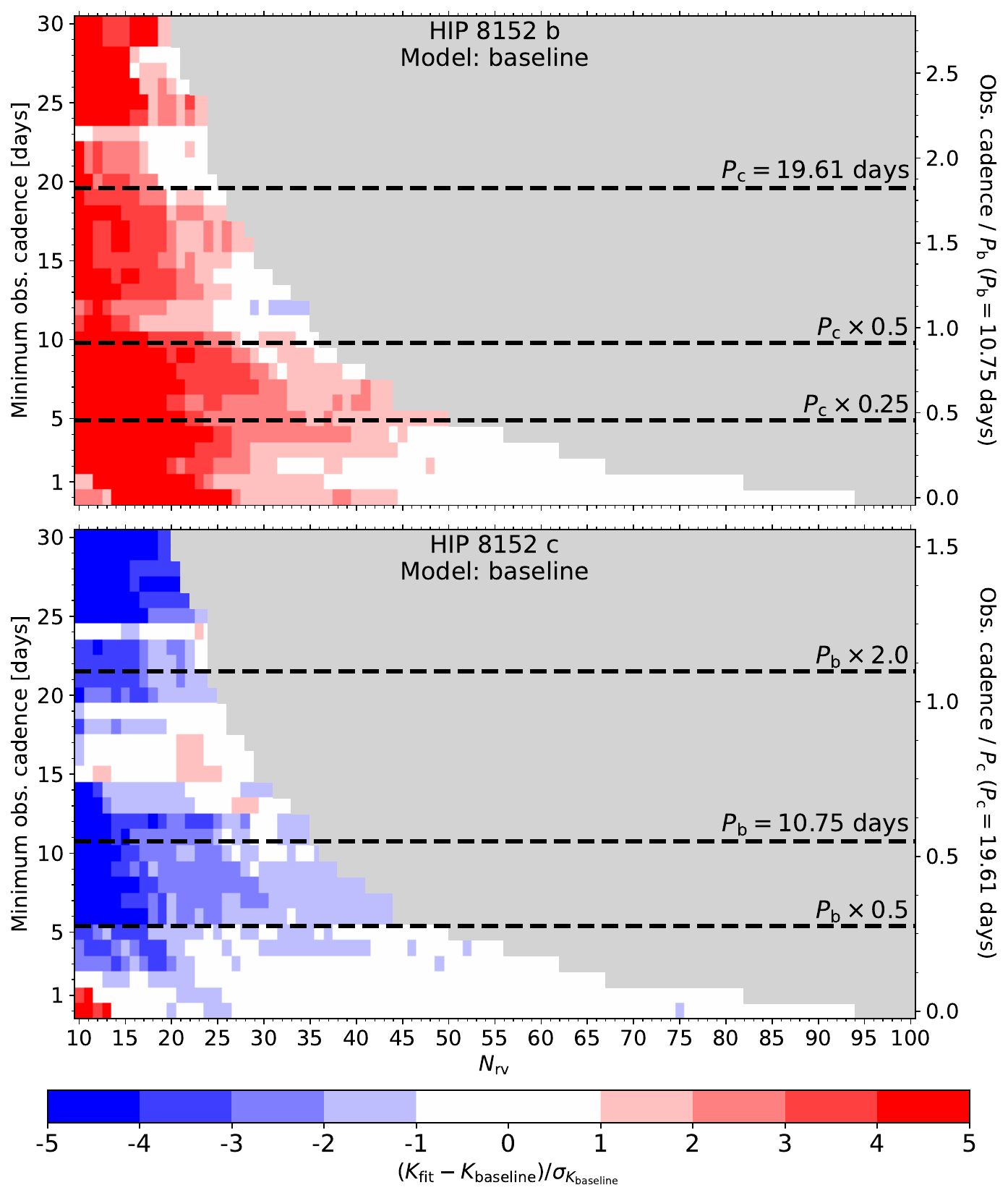}
    \caption{The same as Figure \ref{fig:t000669_real_base} but for \sysIII b (top) and c (bottom). In the top (bottom) panel, black horizontal dashed lines represent harmonics of the orbital period of planet c (b).}
    \label{fig:hip8152_real_base}
\end{figure}

There are two key observations from the results of the \sysIII resampling experiment (Figure \ref{fig:hip8152_real_base}). First, we find that for the same combinations of MOC and \nrv, the mass of planet b is consistently overestimated relative to its baseline value while the mass of planet c is consistently underestimated. Second, the mass of planet c is seemingly more robust to inaccuracy for MOC $\lesssim 5$ compared to planet b. This speaks to what we might intuit a priori about the relationship between MOC and planet orbital period---mass measurement inaccuracy depends on the ratio between observing cadence and planet orbital period. 

\begin{figure}
    \centering
    \includegraphics[width=\columnwidth]{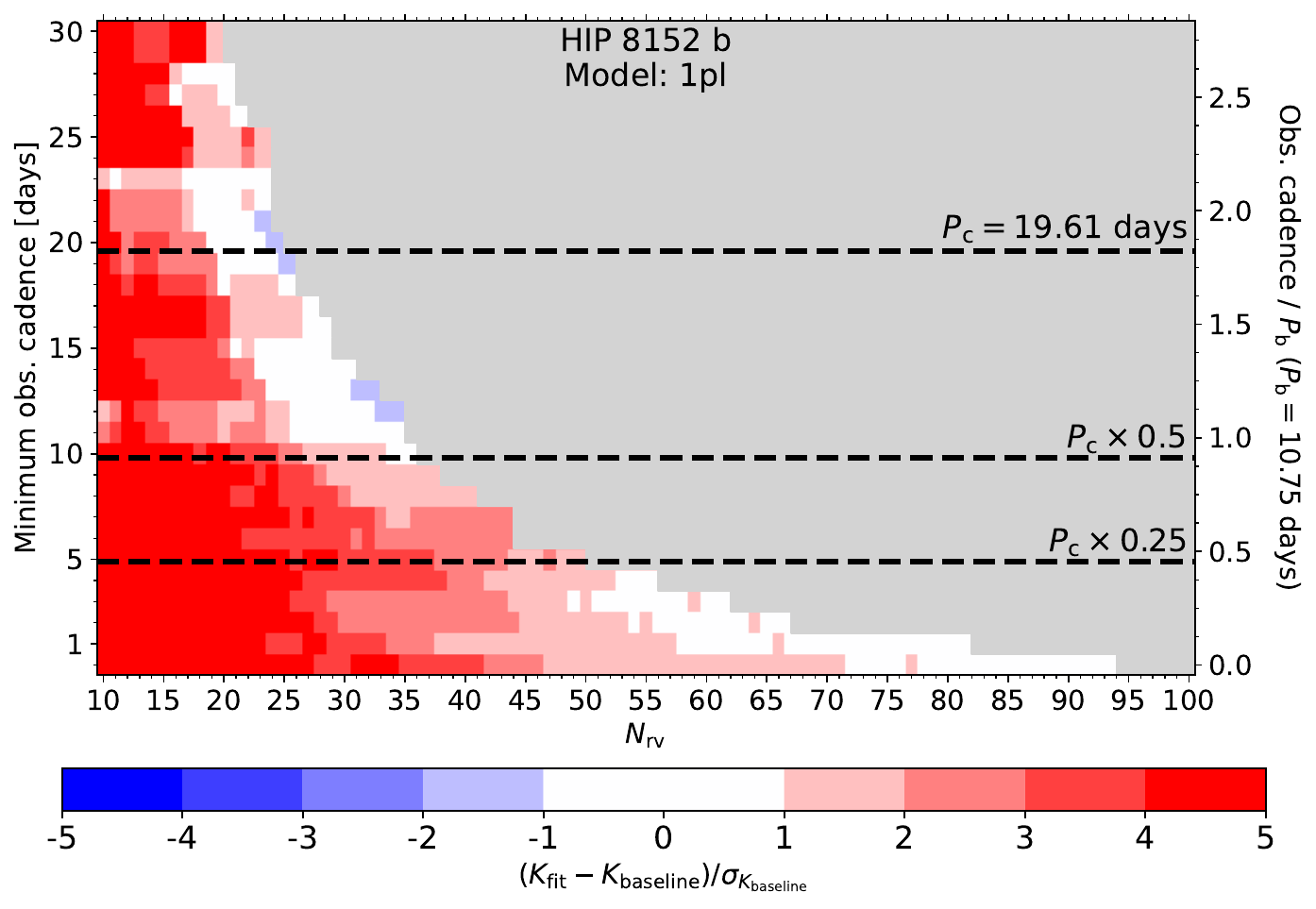}
    \caption{The same as the top panel of Figure \ref{fig:hip8152_real_base}, but instead of using the baseline model we intentionally apply a model that does not account for the orbit of \sysIII c.}
    \label{fig:hip8152_real_1pl}
\end{figure}

In Figure \ref{fig:hip8152_real_1pl}, we show results from the resampling experiment as applied to a model of the \sysIII RVs that does not account for the orbit of \sysIII c. Again, we see that the mass of \sysIII b is systematically overestimated for most combinations of MOC and \nrv, but the issue persists at the 1--2$\sigma$ level even for \nrv $> 50$. In this case, model misspecification leads to an inaccurate planet mass measurement even in the presence of large \nrv. 

\subsection{\sysIV: One transiting planet and one nontransiting planet} \label{sec:toi1444}
\sysIV is an inactive G dwarf known to host a transiting, super-Earth-sized, ultra-short period (USP) planet interior to a nontransiting, sub-Neptune-mass planet candidate. According to \cite{dai21}, the USP, \sysIV b, has $P = 0.47$ days, \rplanet $=$ \rpIVb \rearth, and \mplanet $=$ \mpIVb \mearth, and the nontransiting planet candidate, \sysIV c, has $P = 16.07$ days and $M_\mathrm{p} \sin i_\mathrm{p} =$ \mpIVc \mearth. The orbits of \sysIV b and c are both consistent with $e = 0$. \cite{polanski24} add 22 additional \keckhires RVs to the time series from \cite{dai21}. According to \cite{polanski24}, planet b and c have RV semi-amplitudes of $3.02 \pm 0.62$ \mps and $2.67 \pm 0.72$ \mps, respectively. The full \keckhires data set comprises 74 RVs spanning a baseline of 1306 days.

\begin{figure}
    \centering
    \includegraphics[width=\columnwidth]{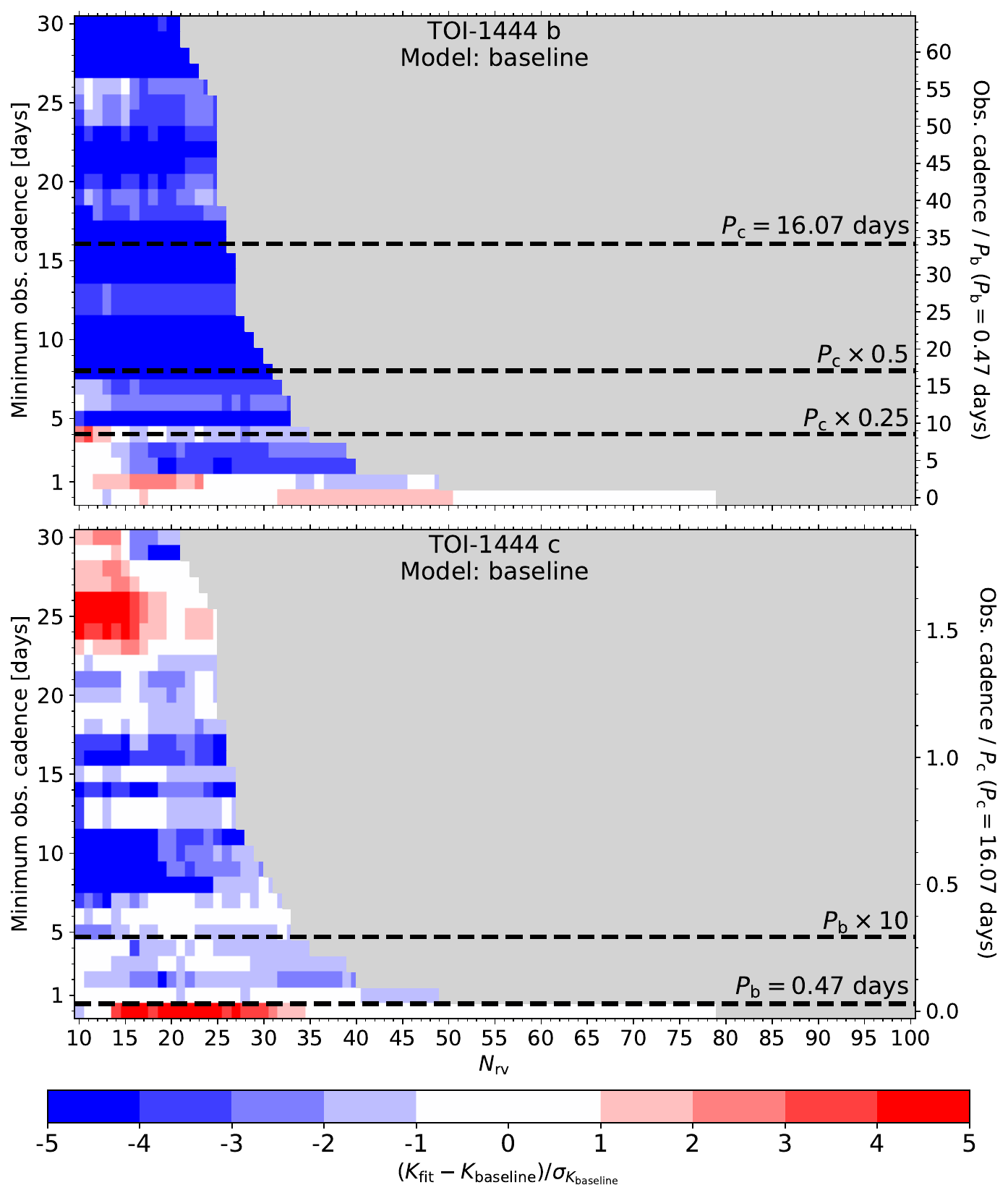}
    \caption{The same as Figure \ref{fig:t000669_real_base} but for \sysIV b (top) and c (bottom). In the top (bottom) panel, black horizontal dashed lines represent harmonics (aliases) of the orbital period of planet c (b).}
    \label{fig:t001444_real_base}
\end{figure}

As expected, the results of the resampling experiment (Figure \ref{fig:t001444_real_base}) show a steep fall off in the agreement between $K_\mathrm{fit}$ and $K_\mathrm{baseline}$ for the USP planet as we downsample the data. USP planets require multiple observations per night for several nights in order to obtain a mass. For the longer-period, nontransiting planet candidate, the agreement also degrades, but not as severely as for planet b. This is similar to the case of \sysIII c, where the accuracy of $K_\mathrm{fit}$ with respect to $K_\mathrm{baseline}$ does not fall off as quickly in MOC as it does for the shorter-period orbit of \sysIII b.

\subsection{\sysV: One transiting planet and one nontransiting planet with an active host star} \label{sec:toi1272}
\sysV is an active G dwarf (\logrhk $=$ \logrhkV) known to host a transiting sub-Neptune and a nontransiting, planetary-mass companion. According to \cite{macdougall22}, \sysV b has $P = 3.32$ days, $e = $ \eVb, \rplanet $=$ \rpVb \rearth, and \mplanet $=$ \mpVb \mearth. \sysV c has $P = 8.69$ days, $e \lesssim 0.35$ (which was constrained using dynamical stability simulations, not the RVs), and $M_\mathrm{p} \sin i_\mathrm{p} =$ \mpVc \mearth. \cite{polanski24} add 14 additional \keckhires RVs to the time series from \cite{macdougall21}. According to \cite{polanski24}, planet b and c have RV semi-amplitudes of $13.5 \pm 1.3$ \mps and $7.4 \pm 1.2$ \mps, respectively. The full \keckhires data set comprises 75 RVs spanning a baseline of 1235 days.

\begin{figure}
    \centering
    \includegraphics[width=\columnwidth]{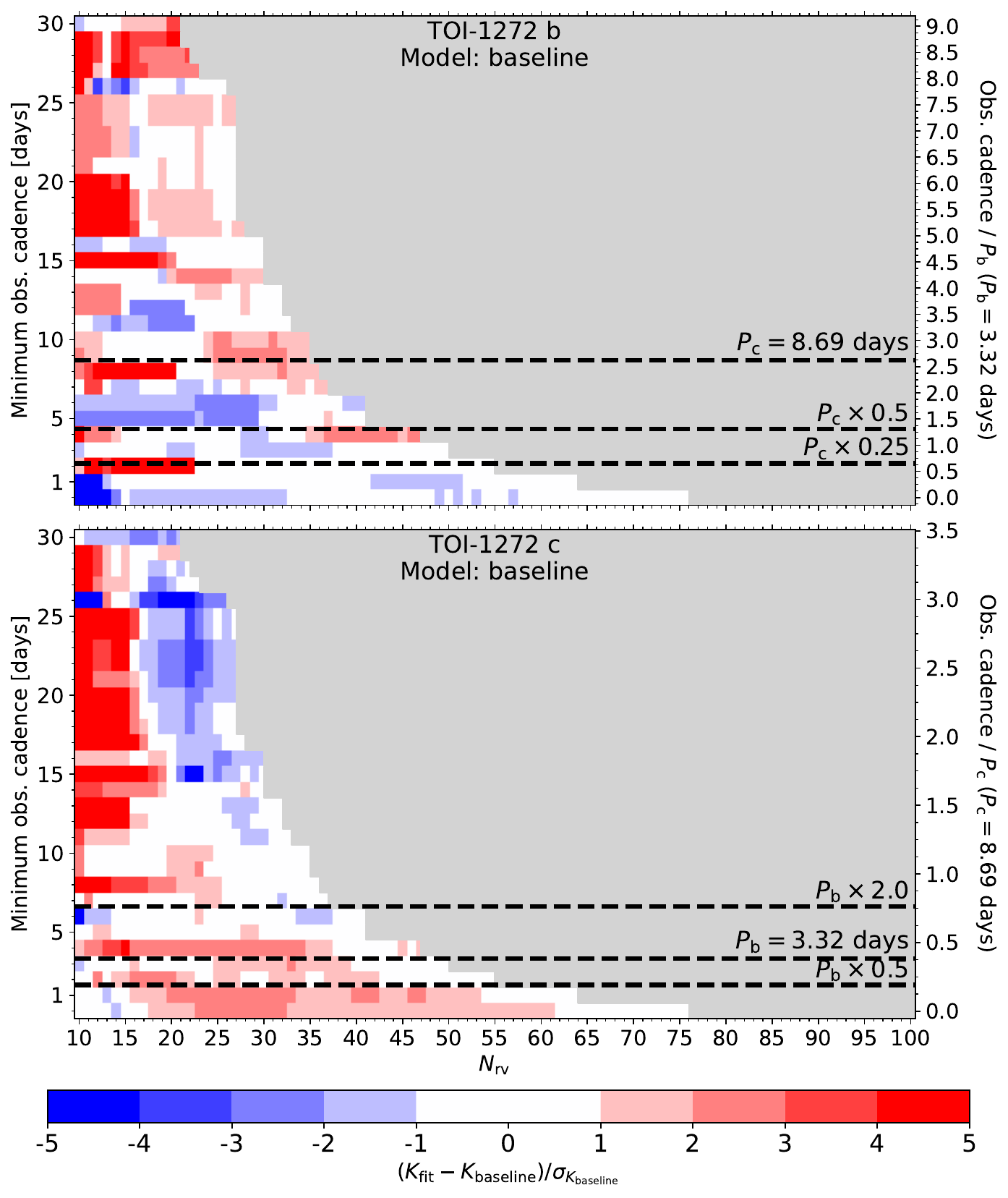}
    \caption{The same as Figure \ref{fig:hip8152_real_base} but for \sysV b (top) and c (bottom). In the top (bottom) panel, black horizontal dashed lines represent harmonics of the orbital period of planet c (b).}
    \label{fig:t001272_real_base}
\end{figure}

The results of our resampling experiment of the \sysV data using our baseline, Keplerian-only model are shown in Figure \ref{fig:t001272_real_base}. Unlike the cases of \sysIII and \sysIV, here we see that the longer-period planet, \sysV c, is seemingly more susceptible to mass measurement inaccuracy than planet b for MOC of a few days or less. For MOC $< 5$ days, we also see that planet b's mass tends to be underestimated and planet c's mass tends to be overestimated, which is opposite compared to \sysIII b and c. 

\begin{figure}
    \centering
    \includegraphics[width=\columnwidth]{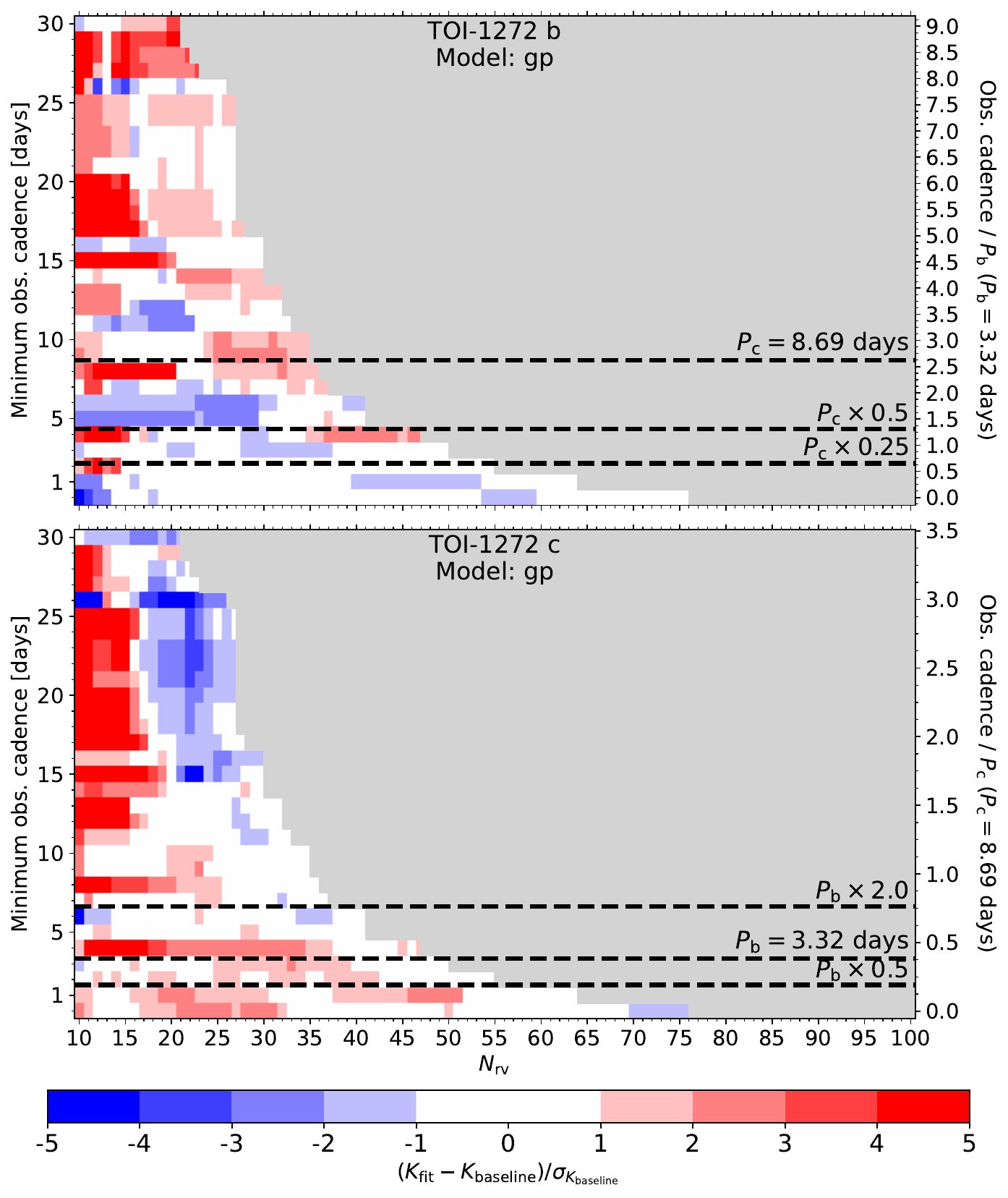}
    \caption{The same as Figure \ref{fig:t001272_real_base} but, but now a GP has been included to account for correlated noise due to stellar activity.}
    \label{fig:t001272_real_gp}
\end{figure}

Figure \ref{fig:t001272_real_gp} shows the results of the resampling experiment where the ``baseline'' model has been augmented with a GP component to account for correlated noise due to stellar activity. The GP uses a quasi-periodic kernel \citep[e.g.,][]{grunblatt15, kosiarek21} which quantifies covariance between data observed as times $t$ and $t'$ as 
\begin{equation} \label{eqn:qp_kernel}
    k (t,t') = \eta_{1}^2 \ \mathrm{exp} \left[-\frac{(t-t')^2}{\eta_2^2}-\frac{\sin^2(\frac{\pi(t-t')}{\eta_3})}{2 \eta_4^2}\right].
\end{equation}
$\eta_{1}$ represents the amplitude of the covariance, $\eta_2$ is interpreted as the evolutionary timescale of active stellar regions, $\eta_3$ is interpreted as the stellar rotation period, and $\eta_4$ is the length scale of the covariance's periodicity.

For planet b, the results of the resampling experiment for the GP-enabled model do not show an obvious change from those of the Keplerian-only model. For planet c, the results are also similar to those of the baseline model, but a fit to the full data set disagrees with the baseline model at the 1--2$\sigma$ level. \cite{macdougall22} suggest $P_\mathrm{rot} = 28.3$ days for \sysV and identify a sub-significant peak in a Lomb-Scargle periodogram of the RV residuals at $P = 14.1$ days (i.e., $P_\mathrm{rot}/2$). Perhaps the measured mass of \sysV c ($P = 8.69$ days) is more dependent on the choice of the adopted model (Keplerian-only or GP-enabled), because it is in the neighborhood of $P_\mathrm{rot}/2$.

\subsection{Takeaways} \label{sec:takeaways}
From our resampling experiments, we find that as observing cadence becomes more infrequent and \nrv decreases, a planet's measured RV semi-amplitude tends to disagree with the value that would otherwise be measured from a fit to the full data set. While one might be tempted to think this result is by design, it is not necessarily obvious a priori that downsampling RV data biases a planet's recovered semi-amplitude---using $\sqrt{N}$-statistics, one might predict a decrease in the precision on $K$ based on the number of data included in the fit, but there is nothing to be said about how the \emph{value} of $K$ itself might change, let alone how it might change with respect to the minimum observing cadence.

In general, we tend to see $\gtrsim2$$\sigma$ disagreement between $K_\mathrm{fit}$ and $K_\mathrm{baseline}$ for MOC $\gtrsim P/3$. These experiments provide empirical evidence to support the idea that inadequate sampling cadence and/or too few data can bias planet mass measurements away from the values that would otherwise be published. Though, from our handful of case studies, it is unclear in which direction (i.e., over- or underestimated) the mass will be biased. We offer two recommendations that observers can follow to help mitigate mass measurement inaccuracy:
\begin{itemize}
    \item Obtain 2--3 RVs per orbital period of the innermost planet in the system.
    \item Plan to acquire \emph{at least} 40 RVs at adequate observing cadence.
\end{itemize}

We note that in the case of USP planets, the sampling cadence requirement is multiple observations per night. We also note that the threshold of $\sim$40 observations agrees with the results of \cite{cochran96}, who found that, for a planet whose RV semi-amplitude is equal to the pointwise measurement error, about 45 independent observations are required to achieve a False Alarm Probability (FAP) of less than 1\%. While these recommendations are not a cure-all, they provide an empirical basis with which to justify the observing strategies and time requests of future Doppler observing proposals.

\section{The impact of undetected companions} \label{sec:synth_data}
Our resampling experiments on real \keckhires data show that inadequate time sampling and/or too few observations can bias planet mass measurements. Additionally, as we saw in the case of \sysIII, applying the incorrect RV model can also cause mass measurement inaccuracy (Figure \ref{fig:hip8152_real_1pl}). In general, though, we do not know what is the ``correct'' RV model, since additional planets may go undetected in photometry and/or RV time series due to small signal sizes, inadequate observing baselines, noise processes, or some combination thereof. The implicit assumption, then, is that the error reported on a planet's mass measurement is large enough to encapsulate this inherent systematic uncertainty.

Is this assumption valid? One might think that if the signal of, for example, a low-mass, wide-orbit planet is too small to detect, then leaving it unaccounted for in a model of the system's RVs should not affect the mass measurement of an inner, transiting planet. While this intuition seems reasonable, we conducted a series of injection-recovery tests using synthetic data to quantify the effect that undetected companions have on the accuracy of the mass measurement of a known planet.

To conduct the simulations, we assumed that a host system with \mstar $= 0.8$ \msun has one ``known'' planet on a circular orbit with a given orbital period and time of inferior conjunction. We generated synthetic RVs using chosen values of $P$ and \mplanet for the known planet, as well as values for the precision on each measurement ($\sigma_\mathrm{RV,\,inst}$) and additional white noise added to each RV ($\sigma_\mathrm{RV,\,astro}$) meant to represent e.g., stellar granulation. The data are generated following the recommendations from \S\ref{sec:takeaways}. We used \radvel to fit the synthetic data and produce a baseline value for the measurement of the known planet's RV semi-amplitude ($K_\mathrm{baseline}$; see Table \ref{tab:radvel_model_synth_data}). Using injection-recovery tests, we then computed completeness contours for each of the synthetic data sets. For each of the injected signals that failed to trigger a detection, we added its Keplerian to the synthetic RVs and refit the augmented data with a one-planet model that only accounted for the orbit of the known planet. The semi-amplitude of the known planet ($K_\mathrm{fit}$) as measured from the augmented data was then compared to $K_\mathrm{baseline}$. We describe the details of our injection-recovery simulations in \S\ref{sec:injection-recovery}.

\input{radvel_model_synth_data.tex}

\subsection{Generating the synthetic RV data}
We simulated four different cases of noise properties using combinations of $\sigma_\mathrm{RV,\,inst} \in [0.4, 1.2]$ \mps and $\sigma_\mathrm{RV,\,astro} \in [0.4, 1.2]$ \mps. For $\sigma_\mathrm{RV,\,inst}$, 0.4 \mps and 1.2 \mps correspond to the internal precision for instruments like the Keck Planet Finder \citep[KPF;][]{gibson16} and \keckhires, respectively. For each noise scenario, we generated synthetic RVs using the four combinations of $P \in [3, 31]$ days and $M_\mathrm{p} \in [2, 20]$ \mearth for the known planet. \transitTime for the planet was chosen randomly from $\mathcal{U}[0, P]$. Following the recommendations from \S\ref{sec:takeaways}, for each of the 16 noise and known planet scenarios, we generated $N_\mathrm{RV} = 40$ synthetic RVs at an observing cadence of $P$/3 (i.e., we generate three RVs per orbit of the known planet). The synthetic RVs were generated as follows. 

First, we chose a list of timestamps to act as the dates of the RV observations. The timestamps were chosen by laying down a grid of times at the exact observing cadence and then adding random noise to each grid point by drawing samples from $\mathcal{U}[-0.33, 0.33]$ days. Initially, the number of grid points generated was equal to 1.3 $\times$ $N_\mathrm{RV}$, then $N_\mathrm{RV} \times 0.3$ of the timestamps were randomly removed to simulate weather losses or telescope scheduling constraints, leaving us with $N_\mathrm{RV}$ timestamps.

With the observation timestamps in hand, we used \radvel to generate the RV orbit of the known planet given the inputs $P$, \transitTime, $e \equiv 0$, and \mstar $= 0.8$ \msun. Random noise was added to each of the RVs using a Gaussian of width $(\sigma_\mathrm{RV,\,inst}^2 + \sigma_\mathrm{RV,\,astro}^2)^{1/2}$. Each synthetic RV was assigned a measurement uncertainty of $\sigma_\mathrm{RV,\,inst}$. Using \radvel, we then fit the synthetic data using the model described in Table \ref{tab:radvel_model_synth_data} and estimated the posterior distribution of the known planet's semi-amplitude to produce $K_\mathrm{baseline}$ and $\sigma_{K_\mathrm{baseline}}$.

In an additional experiment using $\sigma_\mathrm{RV,\,inst} = 0.4$ \mps and $\sigma_\mathrm{RV,\,astro} = 0.4$ \mps, we also added red noise to the synthetic RVs to simulate the impact of untreated stellar activity. The red noise was generated by evaluating the joint prior distribution of a GP at each of the timestamps \citep[see Equation 2.17 in][]{rasmussen06}. The GP used to generate the red noise component had a quasi-periodic kernel (Equation \ref{eqn:qp_kernel}). We chose hyperparameter values of $\eta_1 = 2$ \mps, $\eta_2 = 35$ days, $\eta_3 = 27$ days, and $\eta_4 = 0.5$, which are representative of such values for the Sun \citep{kosiarek20}.

\subsection{Injection-recovery tests} \label{sec:injection-recovery}
With the synthetic RVs and reference values for $K_\mathrm{baseline}$ and $\sigma_{K_\mathrm{baseline}}$ for each of the simulation's configurations, we conducted injection-recovery tests using \rvsearch \citep{rosenthal21}. For each of the 16 combinations of $\sigma_\mathrm{RV,\,inst}$, $\sigma_\mathrm{RV,\,astro}$, $P$, and \mplanet (and for the four additional cases that included the red noise component), we ran 3,000 injection-recovery tests. Pairs of $K$ and $P$ for the injected signals were randomly drawn from $\log_\mathrm{10} K \sim \mathcal{U}[-1, 3]$ \mps and $\log_\mathrm{10} P \sim \mathcal{U}[0, 3]$ days. All of the orbits of the injected signals were forced to be circular.

To determine whether or not the additional signal was ``detected,'' we fit the augmented data with a one- and two-Keplerian model, with orbits corresponding to the known planet and the added companion. We compared then compared the Bayesian Information Criterion \citep[BIC;][]{schwarz78} values of each model. The BIC is defined as 
\begin{equation} \label{eqn:bic}
    \mathrm{BIC} = N_\mathrm{par} \ln N_\mathrm{obs} - 2 \ln \mathcal{\hat{L}}.
\end{equation}
where notation follows as before. Using the detection methodology from \cite{howard16}, we then estimated the $\Delta$BIC value corresponding to a FAP threshold of 0.1\%. If the $\Delta$BIC between the one- and two-Keplerian model corresponded to a detection of the second signal with FAP $< 0.1\%$ threshold, we considered the additional signal to be successfully recovered.

After conducting the injection-recovery tests, we identified all of the injected signals that failed to meet the 0.1\% FAP detection threshold. Upon inspection, some of the signals that went ``undetected'' by \rvsearch were in fact clearly obvious by eye in the synthetic time series, but failed to trigger a detection due to numerical errors in the model-fitting optimization routine. These instances represent a limitation of the data fitting and detection methodology. To largely avoid incorporating these cases in our analysis moving forward, we restricted ourselves to examining only those injected signals within the range $K \in [0.1, 10 \times \mathrm{RMS(RV)}]$ \mps and $P \in [1, 4 \times \tau]$ days, where $\mathrm{RMS(RV)}$ is the raw RMS of the original synthetic time series (containing only noise and the known planet) and $\tau$ is the observing baseline of the synthetic data.

\subsection{The impact of lurking signals}
For all of the simulation scenarios, for each of the $K$ and $P$ pairs that failed to register a detection by \rvsearch, we added the corresponding Keplerian to the synthetic RV time series and fit the augmented data using a one-planet model. We compared the resulting MAP estimate for the RV semi-amplitude of the known planet with $K_\mathrm{baseline}$. 

\begin{figure*}
    \centering
    \includegraphics[width=\textwidth]{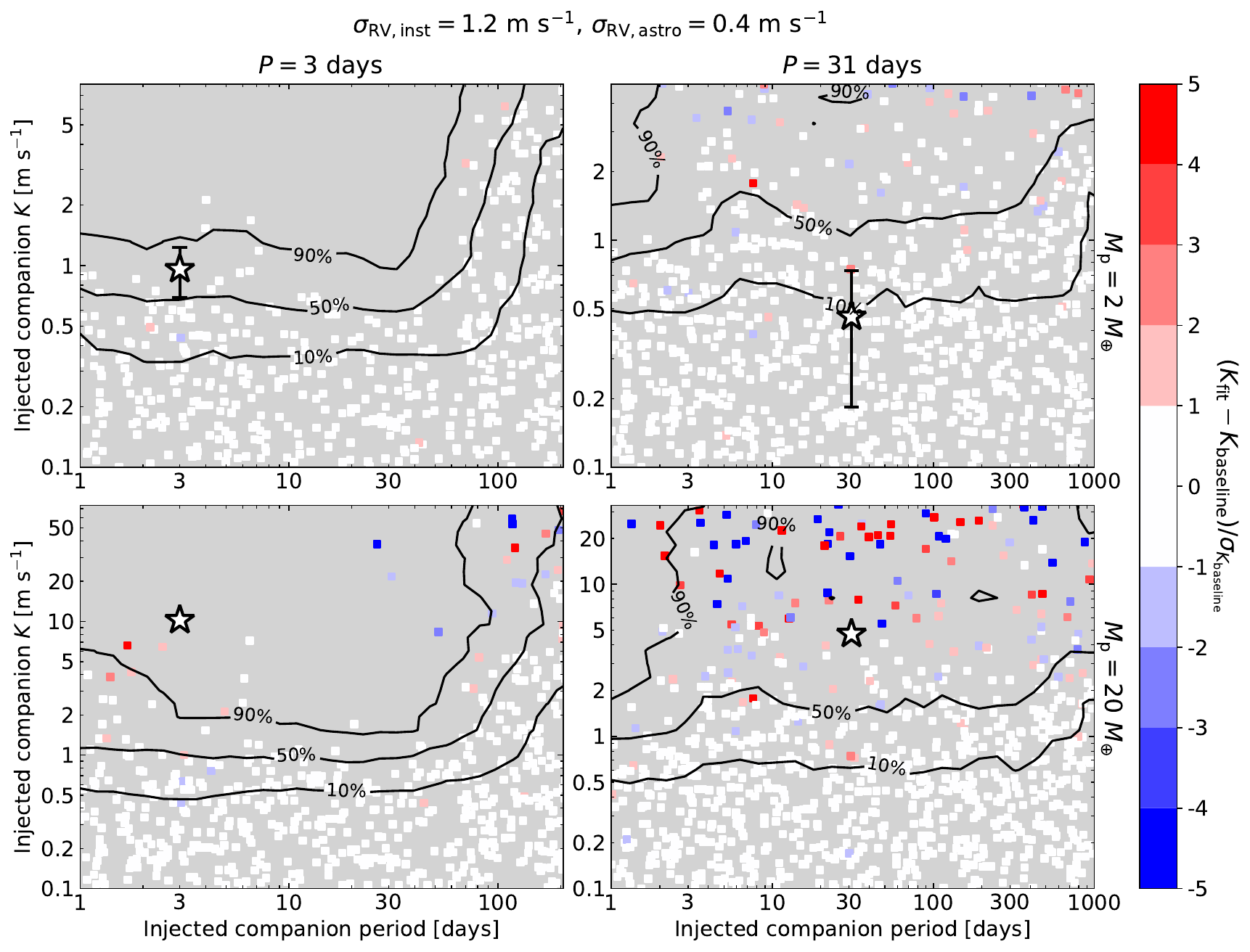}
    \caption{The results of our injection-recovery experiments on the synthetic data in the case of $\sigma_\mathrm{RV,\,inst} = 1.2$ \mps (i.e., the ``HIRES'' case for internal RV precision) and $\sigma_\mathrm{RV,\,astro} = 0.4$ \mps (i.e., the ``quiet'' case for stellar jitter). Each panel represents a combination of $P$ (columns) and \mplanet (rows) for the known planet, which is shown as the white star plotted at $P = P_\mathrm{known}$ and $K = K_\mathrm{baseline}$, and whose the error bar (which is smaller than the size of the marker see in some cases) represents $\sigma_{K_\mathrm{baseline}}$. Filled squares represent individual injected companions that failed to trigger a detection from \rvsearch. The color of each square represents the difference between $K_\mathrm{fit}$ and $K_\mathrm{baseline}$ for the known planet from a one-planet fit to the synthetic RVs that now contains the signal of the undetected companion. The contours represent loci of 90\%, 50\%, and 10\% detection efficiency.}
    \label{fig:four_panel_hires_quiet}
\end{figure*}

\begin{figure*}
    \centering
    \includegraphics[width=\textwidth]{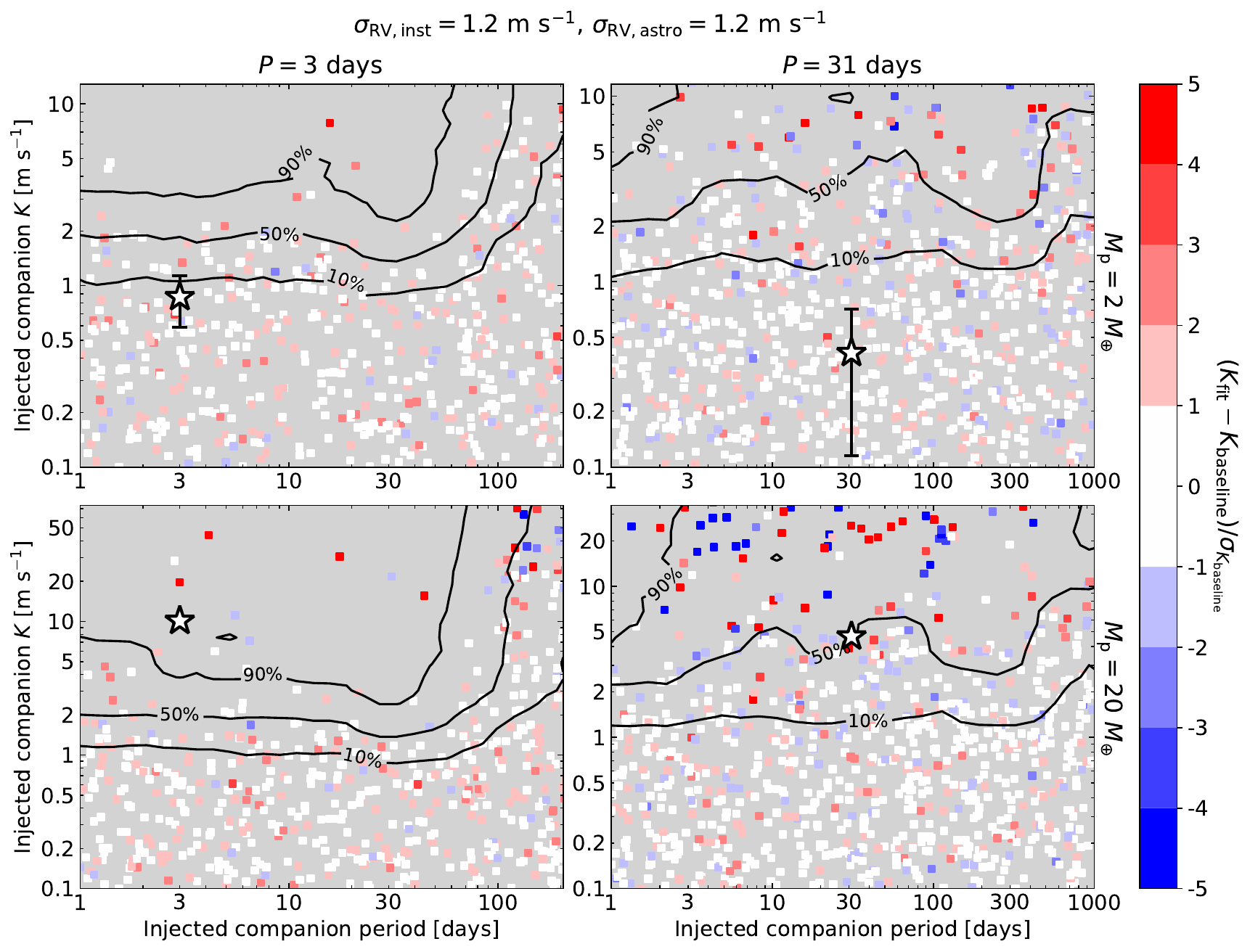}
    \caption{The same as Figure \ref{fig:four_panel_hires_quiet} but now $\sigma_\mathrm{RV,\,astro} = 1.2$ \mps (i.e., the ``active'' case for stellar jitter).}
    \label{fig:four_panel_hires_active}
\end{figure*}

\begin{figure*}
    \centering
    \includegraphics[width=\textwidth]{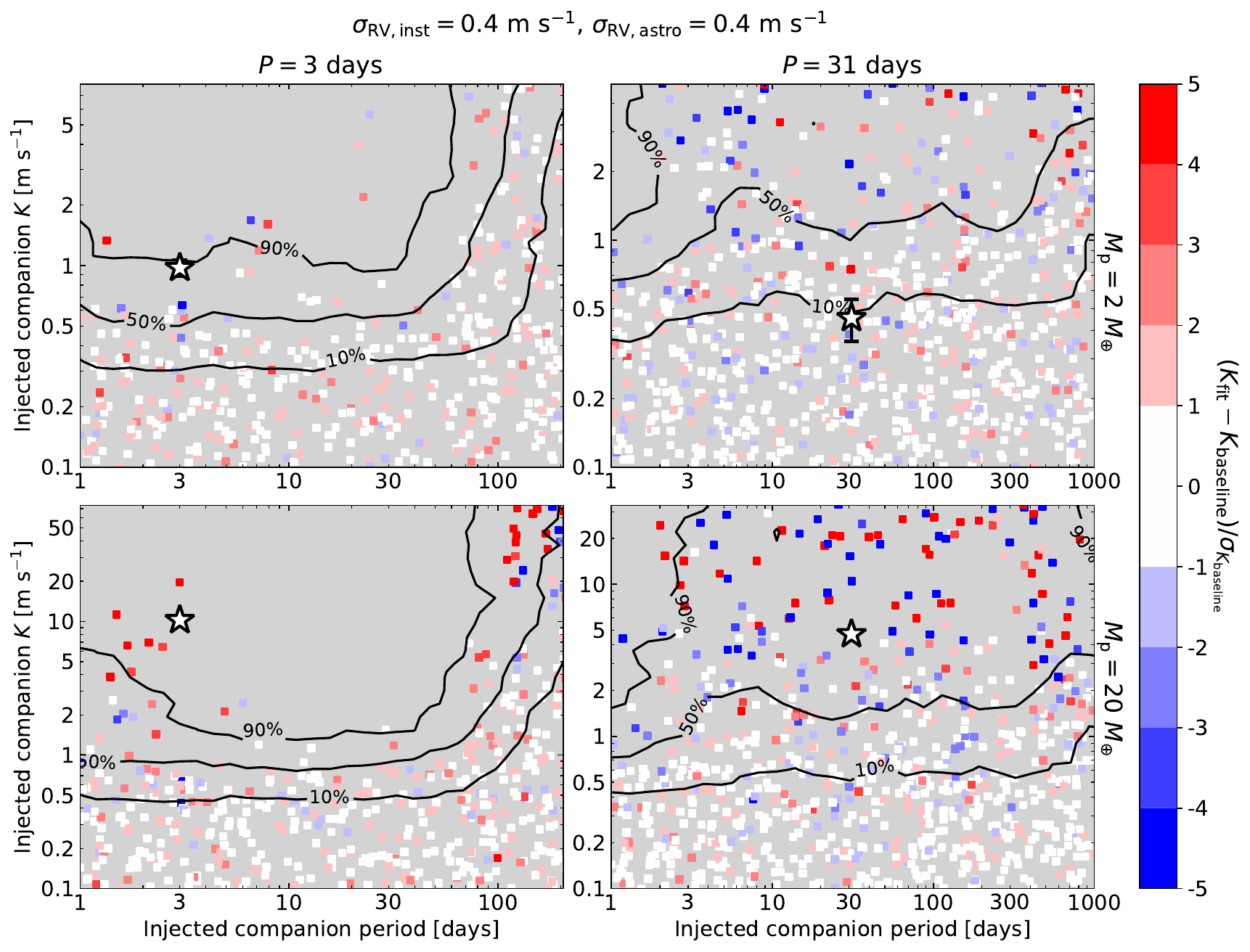}
    \caption{The same as Figure \ref{fig:four_panel_hires_quiet} but now $\sigma_\mathrm{RV,\,inst} = 0.4$ \mps (i.e., the ``KPF'' case for internal RV precision).}
    \label{fig:four_panel_kpf_quiet}
\end{figure*}

\begin{figure*}
    \centering
    \includegraphics[width=\textwidth]{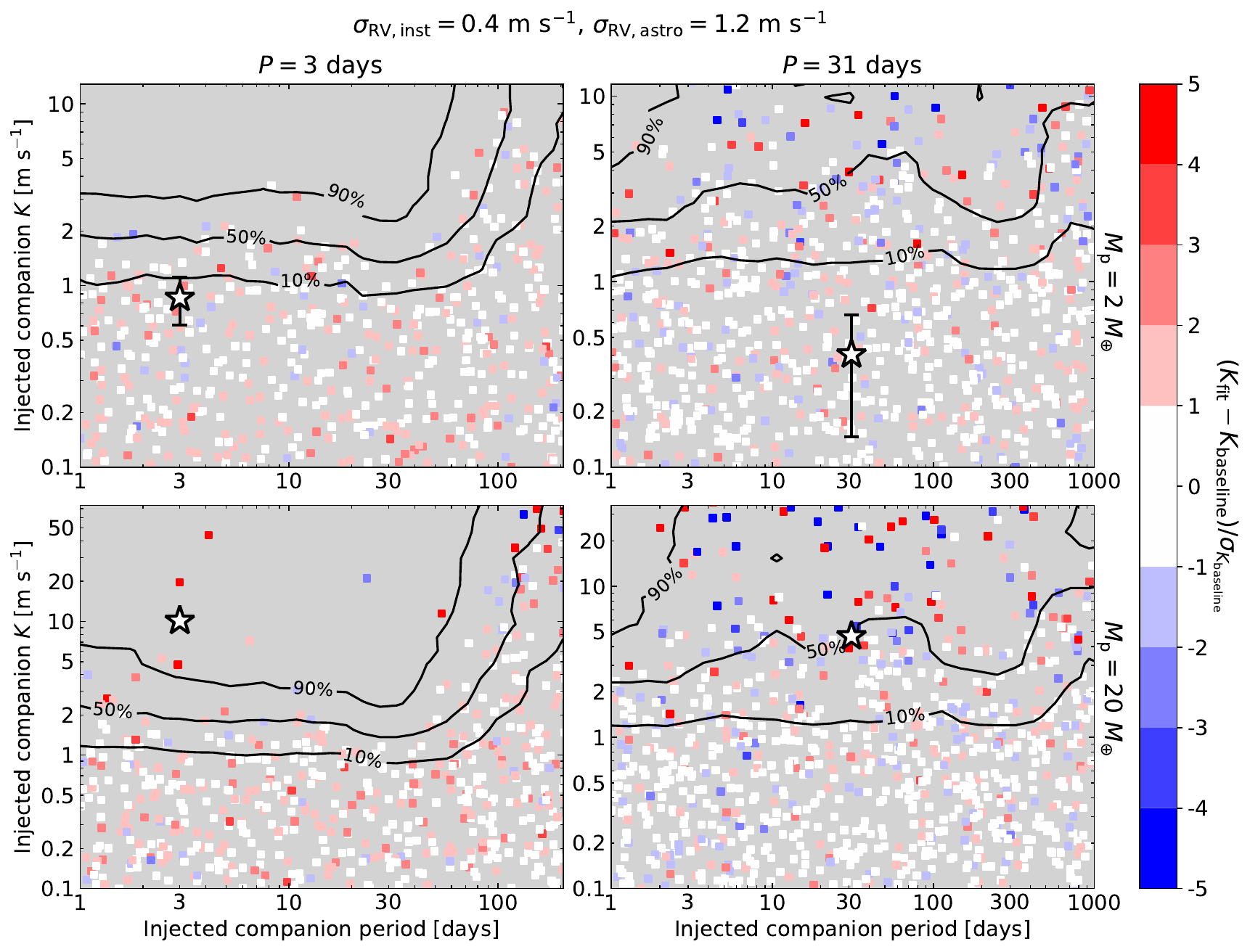}
    \caption{The same as Figure \ref{fig:four_panel_hires_quiet} but now $\sigma_\mathrm{RV,\,inst} = 0.4$ \mps (i.e., the ``KPF'' case for internal RV precision) and $\sigma_\mathrm{RV,\,astro} = 1.2$ \mps (i.e., the ``active'' case for stellar jitter).}
    \label{fig:four_panel_kpf_active}
\end{figure*}

\begin{figure*}
    \centering
    \includegraphics[width=\textwidth]{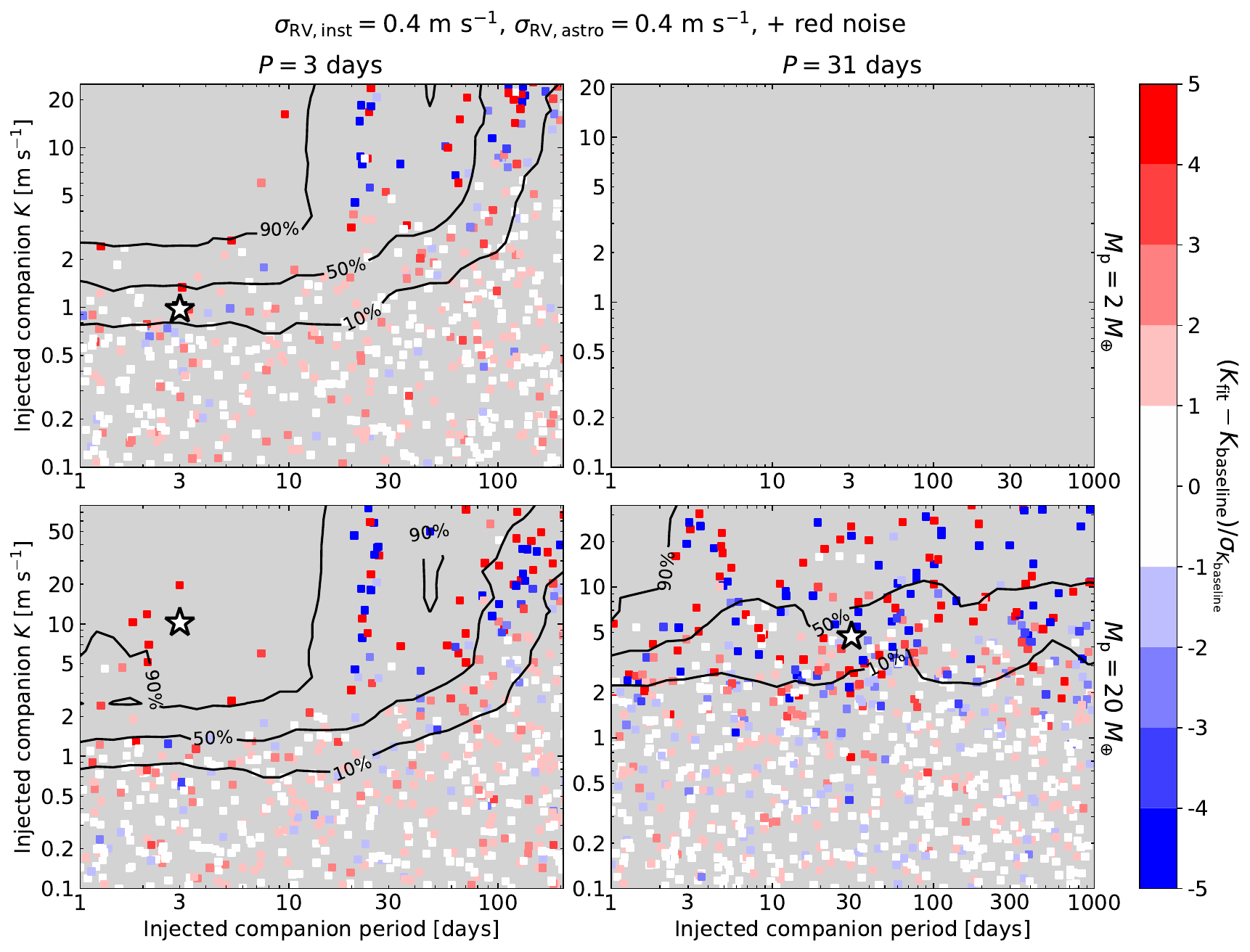}
    \caption{The same as Figure \ref{fig:four_panel_kpf_quiet}, but now the realization of a GP has been added to the synthetic RV time series to mimic correlated noise due to stellar activity. Results for the $P = 31$ days and \mplanet $= 2$ \mearth scenario are not shown due to numerical errors during the RV fitting routine.}
    \label{fig:four_panel_kpf_quiet_red_noise}
\end{figure*}

The results of the experiments are shown in Figures \ref{fig:four_panel_hires_quiet}--\ref{fig:four_panel_kpf_quiet_red_noise}. The following are a few observations. 
\begin{itemize}
    \item For a given value of $\sigma_\mathrm{RV,\:inst}$, disagreement between $K_\mathrm{fit}$ and $K_\mathrm{baseline}$ increases as $\sigma_\mathrm{RV,\,astro}$ increases (compare Figures \ref{fig:four_panel_hires_quiet} and \ref{fig:four_panel_hires_active}). 
    \item In the case of $\sigma_\mathrm{RV,\,astro} = 0.4$ \mps, there tends to be more disagreement between $K_\mathrm{fit}$ and $K_\mathrm{baseline}$ for $\sigma_\mathrm{RV,\:inst} = 0.4$ \mps, than for $\sigma_\mathrm{RV,\:inst} = 1.2$ \mps (compare Figures \ref{fig:four_panel_hires_quiet} and \ref{fig:four_panel_kpf_quiet}). This implies that for state-of-the-art Doppler spectrographs with sub-1 \mps internal precision---for which mass measurements of known planets will have systematically smaller uncertainties compared to measurements using data from less stable, perhaps older, instruments---model misspecification, even in the case of small signals, can still bias planet masses at $\gtrsim 2\sigma$ significance.
    \item In the case of $\sigma_\mathrm{RV,\,astro} = 1.2$ \mps, we find similar results for $\sigma_\mathrm{RV,\:inst} = 0.4$ \mps and $\sigma_\mathrm{RV,\:inst} = 1.2$ \mps (compare Figures \ref{fig:four_panel_hires_active} and \ref{fig:four_panel_kpf_active}), suggesting that even with higher instrumental precision, stellar activity may still limit planet mass measurement accuracy.
    \item Untreated correlated noise (representative of stellar activity) further degrades the agreement between $K_\mathrm{fit}$ and $K_\mathrm{baseline}$ (Figure \ref{fig:four_panel_kpf_quiet_red_noise}).
    \item Generally, undetected companions, especially those falling below the 50\% detection threshold, do not impact the accuracy of the mass measurement of the known planet. This, however, may not be true for short-period planets with inadequate observing baselines (as in the case of \sysI b).
\end{itemize}

\begin{figure}
    \centering
    \includegraphics[width=\columnwidth]{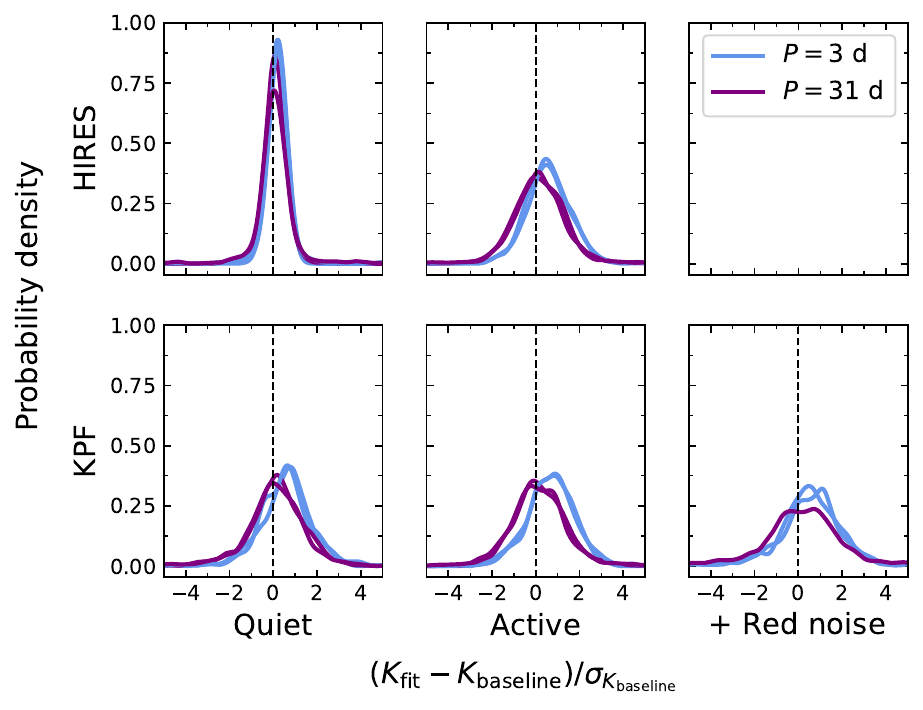}
    \caption{For the different scenarios listed in Table \ref{tab:recovery_test_results}, we show the distributions of the known planet's recovered RV semi-amplitude, $K_\mathrm{fit}$, relative to the baseline value, $K_\mathrm{baseline}$, in units of the uncertainty on the baseline value, $\sigma_{K_\mathrm{baseline}}$. The distributions have been smoothed using a Gaussian kernel with a bandwidth of 0.25 $\sigma_{K_\mathrm{baseline}}$. Blue and purple lines correspond to known planets with $P = 3$ d and $P = 31$ d, respectively. For each orbital period, the lines for the cases of \mplanet $= 2$ \mearth and 20 \mearth generally fall on top of one another, save for the scenario with added red noise.}
    \label{fig:recovery_results_six_panel}
\end{figure}

The results for the simulations shown in Figures \ref{fig:four_panel_hires_quiet}--\ref{fig:four_panel_kpf_quiet_red_noise} are quantified in Table \ref{tab:recovery_test_results}. The marginalized distributions of the failed recoveries' influence on $K_\mathrm{fit}$ are shown in Figure \ref{fig:recovery_results_six_panel}. Interestingly, in all cases of $\sigma_\mathrm{RV,\:inst}$ and $\sigma_\mathrm{RV,\,astro}$, and in the case of the added red noise, we find that irrespective of the known planet's mass, $K_\mathrm{fit}$ is systematically overestimated relative to $K_\mathrm{baseline}$ for known planets with $P = 3$ days. Fitting a Gaussian to each of the distributions in Figure \ref{fig:recovery_results_six_panel}, we find that for all cases where $P = 3$ days, on average, zero corresponds to the $28 \pm 3$ percentile of the distribution. In other words, 72\% of the undetected companions caused a mass overestimate. This is in contrast to the case of $P = 31$ days, where, on average across all experiments, zero corresponds to the $44 \pm 2$ percentile of the distribution. In both cases, the error bar is the RMS about the mean value for the 10 (nine) experiments involving $P = 3$ days ($P = 31$ days).

Perhaps this is a symptom of the shorter observing baseline for the synthetic data in the case of $P = 3$ days, since lurking signals at longer periods will have less of their full orbit sampled, potentially resulting in a systematic shift in the RVs relative to the orbit of the known planet. In the case of $P = 31$ days, the longer observing baseline samples a larger fraction of the corresponding orbital periods for the undetected companions, meaning a companion's time-averaged contribution to the time series should be closer to zero.

\input{recovery_test_results.tex}

\section{Conclusion} \label{sec:conclusion}
In this paper, we investigated the nature of the accuracy of RV mass measurements for small planets. We conducted resampling experiments on public \keckhires RV time series of four precisely-characterized \tess systems to assess the impact of observing cadence and the number of observations on planet mass measurement accuracy (\S\ref{sec:real_data}). We also generated grids of synthetic RV observations and conducted injection-recovery tests to quantify the influence that undetected companions have on the accuracy of mass measurements of planets with known orbits (\S\ref{sec:synth_data}). Our conclusions are summarized as follows:

\begin{itemize}
    \item Degrading RV sampling cadence and/or the number of observations will lead to planet mass measurements that are systematically biased relative to the baseline values that come from fitting the entire data set. 
    
    \item Model misspecification further amplifies the discrepancy. It is difficult to predict, however, in which direction the measurement will be biased (i.e., over- or underestimated). 
    
    \item We recommend that observers plan to acquire 2--3 RVs per orbit of the inner-most planet in the system and at minimum 40 RVs in order to guard against biased mass estimates.

    \item Undetected companions (e.g., additional small planets hiding beneath the RV noise floor) do not generally bias the mass measurements of planets with known orbits. This may not be true in the case of short observing baselines, though.
    
    \item Our resampling experiments provide empirical justification for acquiring high-cadence observations as opposed to low-cadence observations over a longer baseline. The trade-off, however, is that, as we see in Figure \ref{fig:recovery_results_six_panel}, shorter baselines may tend to systematically overestimate short-period planet masses in the presence of undetected companions.
\end{itemize}

While \emph{precise} mass measurements of small planets are essential \citep{batalha19} for follow-up observations with observatories like \jwst and, soon, \ariel \citep{ariel}, observers should also consider the \emph{accuracy} of such masses. Precise yet inaccurate planet masses lead to incorrect conclusions regarding planet bulk and atmospheric composition, as well as formation and evolution scenarios. This is perhaps especially important for the mass measurements of newly-discovered \tess planets, since their shorter observing baselines may not be long enough to fully-characterized the system's architecture.

\vspace{0.5cm}
\begin{center}
ACKNOWLEDGMENTS
\end{center}

The authors thank the anonymous reviewer for their thoughtful report, which improved the manuscript.

All of the code required to reproduce the analysis in this work is available online \citep{murphy24_14029421}.

J.M.A.M. thanks (in alphabetical order) Jacob Bean, Enric Palle, Hannah L. M. Osborne, and María Rosa Zapatero Osorio for helpful discussions and comments that improved the manuscript.

J.M.A.M. is supported by the National Science Foundation (NSF) Graduate Research Fellowship Program (GRFP) under Grant No. DGE-1842400. J.M.A.M. and N.M.B. acknowledge support from NASA’S Interdisciplinary Consortia for Astrobiology Research (NNH19ZDA001N-ICAR) under award number 19-ICAR19\_2-0041.

Support for this work was provided by NASA through the NASA Hubble Fellowship grant \#HF2-51559 awarded by the Space Telescope Science Institute, which is operated by the Association of Universities for Research in Astronomy, Inc., for NASA, under contract NAS5-26555.

This work used \texttt{Expanse} at the San Diego Supercomputer Center through allocation PHY220015 from the Advanced Cyberinfrastructure Coordination Ecosystem: Services \& Support (ACCESS) program, which is supported by NSF grants 2138259, 2138286, 2138307, 2137603, and 2138296.

This work benefited from the 2023 Exoplanet Summer Program in the Other Worlds Laboratory (OWL) at the University of California, Santa Cruz, a program funded by the Heising-Simons Foundation.


\software{\texttt{matplotlib} \citep{matplotlib}, \texttt{numpy} \citep{numpy}, \texttt{pandas} \citep{pandas}, Python 3 \citep{python3}, \radvel \citep{radvel}, \rvsearch \citep{rosenthal21}.}

\bibliography{main}{}
\bibliographystyle{aasjournal}

\end{document}

%% file: author_info.tex
\author[0000-0001-8898-8284]{Joseph M. Akana Murphy}
\altaffiliation{NSF Graduate Research Fellow}
\affiliation{Department of Astronomy and Astrophysics, University of California, Santa Cruz, CA 95064, USA}

\author[0000-0002-4671-2957]{Rafael Luque}
\altaffiliation{NHFP Sagan Fellow}
\affiliation{Department of Astronomy and Astrophysics, University of Chicago, Chicago, IL 60637, USA}

\author[0000-0002-7030-9519]{Natalie M. Batalha}
\affiliation{Department of Astronomy and Astrophysics, University of California, Santa Cruz, CA 95064, USA}

%% file: case_studies_stellar_properties.tex
\begin{deluxetable*}{lccccccccc}
\tablecaption{Properties of host stars in our case studies of real \keckhires data \label{tab:case_studies_stellar_properties}}
\tabletypesize{\small}
\tablehead{\colhead{System name} & \colhead{\teff [K]} & \colhead{[Fe/H] [dex]} & \colhead{\rstar [\rsun]} & \colhead{\mstar [\msun]} & \colhead{\logrhk} & \colhead{$N_\mathrm{Tr}$} & \colhead{$N_\mathrm{NTr}$} & \colhead{$N_\mathrm{RV}$ (binned)} & \colhead{$\sigma_\mathrm{RV}$ [m s$^{-1}$]}}
\startdata
\sysII & $5600 \pm 110$ & $-0.06 \pm 0.09$ & $0.99 \pm 0.04$ & $0.90 \pm 0.06$ & \logrhkII & 1 & 0 & 62 (61) & 1.53 \\
\sysIII & $5530 \pm 110$ & $-0.09 \pm 0.09$ & $0.95 \pm 0.04$ & $0.91 \pm 0.06$ & \logrhkIII & 2 & 0 & 94 (94) & 1.36 \\
\sysIV & $5430 \pm 90$ & $0.13 \pm 0.06$ & $0.91 \pm 0.02$ & $0.93 \pm 0.04$ & \logrhkIV & 1 & 1 & 79 (76) & 1.37 \\
\sysV & $4990 \pm 120$ & $0.17 \pm 0.06$ & $0.79 \pm 0.03$ & $0.85 \pm 0.05$ & \logrhkV & 1 & 1 & 76 (75) & 1.73 \\
\enddata
\tablecomments{
Properties for \sysII and \sysIII come from \cite{murphy23}, properties for \sysIV come from \cite{dai21}, and properties for \sysV come from \cite{macdougall22}. $N_\mathrm{Tr}$ and $N_\mathrm{NTr}$ represent the number of transiting and nontransiting planets assumed in the system's baseline RV model. RVs taken within 0.1 days of each other are binned together during the resampling and fitting procedure. $\sigma_\mathrm{RV}$ represents the median measurement uncertainty of the \keckhires data.
}
\end{deluxetable*}

%% file: radvel_model_real_data.tex
\begin{deluxetable}{lcc}
\tablecaption{\radvel model of resampled data\label{tab:radvel_model_real_data}}
\tabletypesize{\normalsize}
\tablehead{\colhead{Parameter [units]} & \colhead{Prior} & Note}
\startdata
$P$ [days] & $\mathcal{N}$\citepMXXIII & A \\
\transitTime [BJD] & $\mathcal{N}$\citepMXXIII & A\\
$\sqrt{e} \cos \omega_\mathrm{p}$ & $\equiv 0$ & B\\ 
$\sqrt{e} \sin \omega_\mathrm{p}$ & $\equiv 0$ & B \\ 
$K$ [m s$^{-1}$] & $\mathcal{U}$[$-\infty$, $+\infty$] \\
$\gamma_\mathrm{HIRES}$ [m s$^{-1}$] & $\mathcal{U}$[-100, 100] \\
$\sigma_\mathrm{HIRES}$ [m s$^{-1}$] & $\mathcal{U}$[0, 20] \\
\vspace{0.2cm} \\
\multicolumn{2}{l}{\emph{GP hyperparameters (\sysV only)}} & C \\
$\eta_1$ [m s$^{-1}$] & $\mathcal{J}$[0.1, 100] & D\\
$\eta_2$ [days] & $\mathcal{J}$[$\eta_3$, 10000] & D, E \\
$\eta_3$ [days] & $\mathcal{N}$(28.3, 3) & D, F\\
$\eta_4$ & $\mathcal{N}$(0.5, 0.05) & D, G
\enddata
\tablecomments{$\mathcal{J}$[X, Y] refer to a Jeffreys distribution on the interval X and Y \citep{jeffreys46}.\\
\textbf{A}: \citeMXXIII refers to planet ephemerides reported in \cite{macdougall23} as measured from \tess transit photometry. For the nontransiting planets \sysIV c and \sysV c, values for $P$ and \transitTime, as well as their respective errors, were taken from \cite{dai21} and \cite{macdougall22}, respectively. \\
\textbf{B}: Orbital eccentricity is fixed to zero for all Keplerian orbits save for \sysV b \citep{macdougall22}, for which a prior is imposed to keep $e < 0.99$. \\
\textbf{C}: Priors for the GP hyperparameters were chosen to mirror the ones used in \citepolanskiXXIV. \sysV is the only system for which we include a GP component. \\
\textbf{D}: $\eta_1$ is the GP amplitude. $\eta_2$ is the spot evolutionary timescale. $\eta_3$ is the variability's periodic time scale (i.e., the stellar rotation period). $\eta_4$ is the periodic length scale of the variability. \\
\textbf{E}: $\eta_2$ is forced to be larger than $\eta_3$ per \cite{kosiarek20}. \\
\textbf{F}: $P_\mathrm{rot} = 28.3$ days is the stellar rotation period estimate from \cite{macdougall22}, as measured using the star's \texttt{TESS-SIP} light curve \citep{hedges20} and \keckhires \shk values \citep{isaacson10}.\\
\textbf{G}: Informed prior on $\eta_4$ per \cite{haywood18}. \\
}
\end{deluxetable}

%% file: radvel_model_synth_data.tex
\begin{deluxetable}{lc}
\tablecaption{\radvel model of synthetic data\label{tab:radvel_model_synth_data}}
\tabletypesize{\normalsize}
\tablehead{\colhead{Parameter} & \colhead{Prior}}
\startdata
$P$ [days] & $\equiv P_\mathrm{known}$ \\
\transitTime [BJD] & $\equiv T_\mathrm{c,\:known}$ \\
$\sqrt{e}\cos \omega_\mathrm{p}$& $\equiv 0$ \\ 
$\sqrt{e}\sin \omega_\mathrm{p}$ & $\equiv 0$ \\ 
$K$ [m s$^{-1}$] & $\mathcal{U}$[$-\infty$, $+\infty$] \\
$\gamma$ [m s$^{-1}$] & $\mathcal{U}$[-100, 100] \\
$\sigma$ [m s$^{-1}$] & $\mathcal{U}$[0, 20] \\
\enddata
\tablecomments{When fitting the synthetic RVs, the period and time of inferior conjunction for the orbit of the known planet are fixed to their true values, $P_\mathrm{known}$ and $T_\mathrm{c,\:known}$.}
\end{deluxetable}

%% file: recovery_test_results.tex
\begin{deluxetable*}{rcrccc|ccccccc} \label{tab:recovery_test_results}
\tablecaption{Impact of undetected signals on the accuracy of the known planet's recovered semi-amplitude, $K_\mathrm{fit}$.}
\tablehead{
  \colhead{} &
  \colhead{RV precision} &
  \colhead{} &
  \colhead{Astro. jitter} & 
  \colhead{\mplanet} & 
  \colhead{$P$} &
  \colhead{$N_\mathrm{tot}$} & 
  \multicolumn{6}{c}{Fraction of failed recoveries' influence on $K_\mathrm{fit}$} \\
  \colhead{} &
  \colhead{[m s$^{-1}$]} &
  \colhead{} &
  \colhead{[m s$^{-1}$]} & 
  \colhead{[$M_\mathrm{\oplus}$]} & 
  \colhead{[days]} &
  \colhead{} & 
  \colhead{$\leq1$$\sigma$} & 
  \colhead{1--2$\sigma$} & 
  \colhead{2--3$\sigma$} & 
  \colhead{3--4$\sigma$} & 
  \colhead{4--5$\sigma$} & 
  \colhead{$\geq5$$\sigma$}
}
\startdata
\ldelim{\{}{9}{*}[HIRES] & 1.2 & \ldelim{\{}{4}{*}[Quiet] & 0.4 & 2 & 3 & 452 & 0.99 & 0.01 & 0.00 & 0.00 & 0.00 & 0.00 \\
& 1.2 & & 0.4 & 2 & 31 & 791 & 0.95 & 0.04 & 0.01 & 0.00 & 0.00 & 0.00 \\
& 1.2  & & 0.4 & 20 & 3 & 511 & 0.94 & 0.04 & 0.01 & 0.00 & 0.00 & 0.01 \\
& 1.2  & & 0.4 & 20 & 31 & 834 & 0.86 & 0.07 & 0.02 & 0.01 & 0.01 & 0.04 \\
\\ 
& 1.2  & \ldelim{\{}{4}{*}[Active] & 1.2 & 2 & 3 & 521 & 0.64 & 0.29 & 0.07 & 0.00 & 0.00 & 0.00 \\
& 1.2  & & 1.2 & 2 & 31 & 842 & 0.64 & 0.29 & 0.05 & 0.01 & 0.00 & 0.00 \\
& 1.2  & & 1.2 & 20 & 3 & 531 & 0.62 & 0.28 & 0.08 & 0.01 & 0.01 & 0.01 \\
& 1.2  & & 1.2 & 20 & 31 & 861 & 0.63 & 0.24 & 0.06 & 0.02 & 0.01 & 0.04 \\
\\
\ldelim{\{}{9}{*}[KPF] & 0.4 & \ldelim{\{}{4}{*}[Quiet] & 0.4 & 2 & 3 & 434 & 0.61 & 0.30 & 0.08 & 0.01 & 0.00 & 0.00 \\
& 0.4 & & 0.4 & 2 & 31 & 788 & 0.61 & 0.27 & 0.08 & 0.02 & 0.00 & 0.02 \\
& 0.4 & & 0.4 & 20 & 3 & 455 & 0.56 & 0.28 & 0.09 & 0.03 & 0.01 & 0.04 \\
& 0.4 & & 0.4 & 20 & 31 & 801 & 0.59 & 0.21 & 0.08 & 0.03 & 0.01 & 0.08 \\
\\
& 0.4 & \ldelim{\{}{4}{*}[Active] & 1.2 & 2 & 3 & 515 & 0.55 & 0.35 & 0.09 & 0.01 & 0.00 & 0.00 \\
& 0.4 & & 1.2 & 2 & 31 & 838 & 0.61 & 0.29 & 0.07 & 0.02 & 0.00 & 0.01 \\
& 0.4 & & 1.2 & 20 & 3 & 545 & 0.54 & 0.33 & 0.09 & 0.02 & 0.01 & 0.01 \\
& 0.4 & & 1.2 & 20 & 31 & 892 & 0.62 & 0.25 & 0.07 & 0.02 & 0.01 & 0.03 \\
\\
\multicolumn{4}{c}{\emph{Plus red noise component}} \\
\ldelim{\{}{4}{*}[KPF] & 0.4 & \ldelim{\{}{4}{*}[Quiet] & 0.4 & 2 & 3 & 553 & 0.54 & 0.25 & 0.12 & 0.03 & 0.01 & 0.05 \\
& 0.4 & & 0.4 & 2 & 31 & \nodata  & \nodata  & \nodata & \nodata & \nodata  & \nodata  & \nodata \\ 
& 0.4 & & 0.4 & 20 & 3 & 556 & 0.44 & 0.32 & 0.11 & 0.04 & 0.02 & 0.08 \\
& 0.4 & & 0.4 & 20 & 31 & 897 & 0.46 & 0.26 & 0.10 & 0.04 & 0.03 & 0.11 \\
\enddata
\tablecomments{\textit{Left side:} Experiment parameters. RV precision refers to the measurement error associated with each synthetic data point. The two fiducial cases we explored, 1.2 \mps and 0.4 \mps precision, correspond to a HIRES-like and a KPF-like instrument, respectively. Astrophysical jitter corresponds to the width of the Gaussian used to add white noise to the synthetic data. This added noise is used as a proxy for astrophysical jitter from e.g., stellar granulation. The two cases we tested corresponded to adding noise from a Gaussian of width 0.4 \mps (``quiet'') and 1.2 \mps (``active''). \mplanet and $P$ correspond to the mass and orbital period of the ``known'' planet from which the synthetic RVs are generated. In all cases, \mstar $= 0.8$ \msun. \textit{Right side:} Experimental results. For each experiment, $N_\mathrm{tot}$ is the total number of injected planetary signals that we failed to detect within the range $K \in [0.1, 10\times \mathrm{RMS(RV)}]$ \mps and $P \in [1, 4 \times \tau]$ days. The next six columns show the fraction of undetected injected signals that cause the recovered RV semi-amplitude of the known planet ($K_\mathrm{fit}$) to deviate from the baseline value by a certain amount. \textit{Bottom:} The bottom four rows reproduce the KPF-like precision plus low astrophysical jitter scenario, but now the residuals include realizations of a GP meant to mimic correlated stellar activity. In the case of \mplanet $= 2$ \mearth and $P = 31$ days, the GP component dominates over the signal of the known planet, leading to numerical issues for our injection-recovery and MAP fitting algorithm.}
\end{deluxetable*}

%% file: main.bbl
\begin{thebibliography}{}
\expandafter\ifx\csname natexlab\endcsname\relax\def\natexlab#1{#1}\fi
\providecommand{\url}[1]{\href{#1}{#1}}
\providecommand{\dodoi}[1]{doi:~\href{http://doi.org/#1}{\nolinkurl{#1}}}
\providecommand{\doeprint}[1]{\href{http://ascl.net/#1}{\nolinkurl{http://ascl.net/#1}}}
\providecommand{\doarXiv}[1]{\href{https://arxiv.org/abs/#1}{\nolinkurl{https://arxiv.org/abs/#1}}}

\bibitem[{{Adams} {et~al.}(2008){Adams}, {Seager}, \& {Elkins-Tanton}}]{adams08}
{Adams}, E.~R., {Seager}, S., \& {Elkins-Tanton}, L. 2008, \apj, 673, 1160, \dodoi{10.1086/524925}

\bibitem[{{Aguichine} {et~al.}(2021){Aguichine}, {Mousis}, {Deleuil}, \& {Marcq}}]{aguichine21}
{Aguichine}, A., {Mousis}, O., {Deleuil}, M., \& {Marcq}, E. 2021, \apj, 914, 84, \dodoi{10.3847/1538-4357/abfa99}

\bibitem[{{Akaike}(1974)}]{akaike74}
{Akaike}, H. 1974, IEEE Transactions on Automatic Control, 19, 716

\bibitem[{{Akana Murphy}(2024)}]{murphy24_14029421}
{Akana Murphy}, J.~M. 2024, {Simulate RV observations by either resampling real data or creating synthetic data}, 0.1.1,  Zenodo, \dodoi{10.5281/zenodo.14029421}

\bibitem[{{Akana Murphy} {et~al.}(2023){Akana Murphy}, {Batalha}, {Scarsdale}, {Isaacson}, {Ciardi}, {Gonzales}, {Giacalone}, {Twicken}, {Dattilo}, {Fetherolf}, {Rubenzahl}, {Crossfield}, {Dressing}, {Fulton}, {Howard}, {Huber}, {Kane}, {Petigura}, {Robertson}, {Roy}, {Weiss}, {Beard}, {Chontos}, {Dai}, {Rice}, {Van Zandt}, {Lubin}, {Blunt}, {Polanski}, {Behmard}, {Dalba}, {Hill}, {Rosenthal}, {Brinkman}, {Mayo}, {Turtelboom}, {Angelo}, {Mo{\v{c}}nik}, {MacDougall}, {Pidhorodetska}, {Tyler}, {Kosiarek}, {Holcomb}, {Louden}, {Hirsch}, {Gilbert}, {Anderson}, \& {Valenti}}]{murphy23}
{Akana Murphy}, J.~M., {Batalha}, N.~M., {Scarsdale}, N., {et~al.} 2023, \aj, 166, 153, \dodoi{10.3847/1538-3881/ace2ca}

\bibitem[{{Akana Murphy} {et~al.}(2024){Akana Murphy}, {Luque}, {Batalha}, {Carleo}, {Palle}, {Brady}, {Fulton}, {Handley}, {Isaacson}, {Lacedelli}, {Murgas}, {Nowak}, {Orell-Miquel}, {Osborne}, {Van Eylen}, \& {Zapatero Osorio}}]{hd119130murphy24arxiv}
{Akana Murphy}, J.~M., {Luque}, R., {Batalha}, N.~M., {et~al.} 2024, arXiv e-prints, arXiv:2411.02518, \dodoi{10.48550/arXiv.2411.02518}

\bibitem[{{Batalha} {et~al.}(2019){Batalha}, {Lewis}, {Fortney}, {Batalha}, {Kempton}, {Lewis}, \& {Line}}]{batalha19}
{Batalha}, N.~E., {Lewis}, T., {Fortney}, J.~J., {et~al.} 2019, \apjl, 885, L25, \dodoi{10.3847/2041-8213/ab4909}

\bibitem[{{Burnham} \& {Anderson}(2002)}]{burnham02}
{Burnham}, K.~P., \& {Anderson}, D.~R. 2002, Model Selection and Multimodel Inference (Springer-Verlag New York)

\bibitem[{Burnham \& Anderson(2004)}]{burnham04}
Burnham, K.~P., \& Anderson, D.~R. 2004, Sociological Methods \& Research, 33, 261, \dodoi{10.1177/0049124104268644}

\bibitem[{{Burt} {et~al.}(2018){Burt}, {Holden}, {Wolfgang}, \& {Bouma}}]{burt18}
{Burt}, J., {Holden}, B., {Wolfgang}, A., \& {Bouma}, L.~G. 2018, \aj, 156, 255, \dodoi{10.3847/1538-3881/aae697}

\bibitem[{{Burt} {et~al.}(2024){Burt}, {Hooton}, {Mamajek}, {Barrag{\'a}n}, {Millholland}, {Fairnington}, {Fisher}, {Halverson}, {Huang}, {Brady}, {Seifahrt}, {Gaidos}, {Luque}, {Kasper}, \& {Bean}}]{burt24}
{Burt}, J.~A., {Hooton}, M.~J., {Mamajek}, E.~E., {et~al.} 2024, \apjl, 971, L12, \dodoi{10.3847/2041-8213/ad5b52}

\bibitem[{Butler {et~al.}(1996)Butler, Marcy, Williams, McCarthy, Dosanjh, \& Vogt}]{butler96}
Butler, R.~P., Marcy, G.~W., Williams, E., {et~al.} 1996, Publications of the Astronomical Society of the Pacific, 108, 500, \dodoi{10.1086/133755}

\bibitem[{{Cadieux} {et~al.}(2024){Cadieux}, {Plotnykov}, {Doyon}, {Valencia}, {Jahandar}, {Dang}, {Turbet}, {Fauchez}, {Cloutier}, {Cherubim}, {Artigau}, {Cook}, {Edwards}, {Hallatt}, {Charnay}, {Bouchy}, {Allart}, {Mignon}, {Baron}, {Barros}, {Benneke}, {Canto Martins}, {Cowan}, {De Medeiros}, {Delfosse}, {Delgado-Mena}, {Dumusque}, {Ehrenreich}, {Frensch}, {Gonz{\'a}lez Hern{\'a}ndez}, {Hara}, {Lafreni{\`e}re}, {Lo Curto}, {Malo}, {Melo}, {Mounzer}, {Passeger}, {Pepe}, {Poulin-Girard}, {Santos}, {Sosnowska}, {Su{\'a}rez Mascare{\~n}o}, {Thibault}, {Vaulato}, {Wade}, \& {Wildi}}]{cadieux24}
{Cadieux}, C., {Plotnykov}, M., {Doyon}, R., {et~al.} 2024, \apjl, 960, L3, \dodoi{10.3847/2041-8213/ad1691}

\bibitem[{{Chontos} {et~al.}(2022){Chontos}, {Murphy}, {MacDougall}, {Fetherolf}, {Van Zandt}, {Rubenzahl}, {Beard}, {Huber}, {Batalha}, {Crossfield}, {Dressing}, {Fulton}, {Howard}, {Isaacson}, {Kane}, {Petigura}, {Robertson}, {Roy}, {Weiss}, {Behmard}, {Dai}, {Dalba}, {Giacalone}, {Hill}, {Lubin}, {Mayo}, {Mo{\v{c}}nik}, {Polanski}, {Rosenthal}, {Scarsdale}, {Turtelboom}, {Ricker}, {Vanderspek}, {Latham}, {Seager}, {Winn}, {Jenkins}, {Quinn}, {Guerrero}, {Collins}, {Ciardi}, {Shporer}, {Goeke}, {Levine}, {Ting}, {Bieryla}, {Collins}, {Kielkopf}, {Barkaoui}, {Benni}, {Esparza-Borges}, {Conti}, {Hooton}, {Kagetani}, {Laloum}, {Marino}, {Massey}, {Murgas}, {Papini}, {Schwarz}, {Srdoc}, {Stockdale}, {Wang}, {Wittrock}, \& {Zou}}]{chontos22}
{Chontos}, A., {Murphy}, J. M.~A., {MacDougall}, M.~G., {et~al.} 2022, \aj, 163, 297, \dodoi{10.3847/1538-3881/ac6266}

\bibitem[{{Cochran} \& {Hatzes}(1996)}]{cochran96}
{Cochran}, W.~D., \& {Hatzes}, A.~P. 1996, \apss, 241, 43, \dodoi{10.1007/BF00644214}

\bibitem[{{Dai} {et~al.}(2021){Dai}, {Howard}, {Batalha}, {Beard}, {Behmard}, {Blunt}, {Brinkman}, {Chontos}, {Crossfield}, {Dalba}, {Dressing}, {Fulton}, {Giacalone}, {Hill}, {Huber}, {Isaacson}, {Kane}, {Lubin}, {Mayo}, {Mo{\v{c}}nik}, {Akana Murphy}, {Petigura}, {Rice}, {Robertson}, {Rosenthal}, {Roy}, {Rubenzahl}, {Weiss}, {Zandt}, {Beichman}, {Ciardi}, {Collins}, {Gonzales}, {Howell}, {Matson}, {Matthews}, {Schlieder}, {Schwarz}, {Ricker}, {Vanderspek}, {Latham}, {Seager}, {Winn}, {Jenkins}, {Caldwell}, {Colon}, {Dragomir}, {Lund}, {McLean}, {Rudat}, \& {Shporer}}]{dai21}
{Dai}, F., {Howard}, A.~W., {Batalha}, N.~M., {et~al.} 2021, \aj, 162, 62, \dodoi{10.3847/1538-3881/ac02bd}

\bibitem[{{Dumusque} {et~al.}(2014){Dumusque}, {Bonomo}, {Haywood}, {Malavolta}, {S{\'e}gransan}, {Buchhave}, {Collier Cameron}, {Latham}, {Molinari}, {Pepe}, {Udry}, {Charbonneau}, {Cosentino}, {Dressing}, {Figueira}, {Fiorenzano}, {Gettel}, {Harutyunyan}, {Horne}, {Lopez-Morales}, {Lovis}, {Mayor}, {Micela}, {Motalebi}, {Nascimbeni}, {Phillips}, {Piotto}, {Pollacco}, {Queloz}, {Rice}, {Sasselov}, {Sozzetti}, {Szentgyorgyi}, \& {Watson}}]{dumusque14}
{Dumusque}, X., {Bonomo}, A.~S., {Haywood}, R.~D., {et~al.} 2014, \apj, 789, 154, \dodoi{10.1088/0004-637X/789/2/154}

\bibitem[{{Fulton} {et~al.}(2018){Fulton}, {Petigura}, {Blunt}, \& {Sinukoff}}]{radvel}
{Fulton}, B.~J., {Petigura}, E.~A., {Blunt}, S., \& {Sinukoff}, E. 2018, \pasp, 130, 044504, \dodoi{10.1088/1538-3873/aaaaa8}

\bibitem[{{Gibson} {et~al.}(2016){Gibson}, {Howard}, {Marcy}, {Edelstein}, {Wishnow}, \& {Poppett}}]{gibson16}
{Gibson}, S.~R., {Howard}, A.~W., {Marcy}, G.~W., {et~al.} 2016, in Society of Photo-Optical Instrumentation Engineers (SPIE) Conference Series, Vol. 9908, Ground-based and Airborne Instrumentation for Astronomy VI, ed. C.~J. {Evans}, L.~{Simard}, \& H.~{Takami}, 990870, \dodoi{10.1117/12.2233334}

\bibitem[{{Grunblatt} {et~al.}(2015){Grunblatt}, {Howard}, \& {Haywood}}]{grunblatt15}
{Grunblatt}, S.~K., {Howard}, A.~W., \& {Haywood}, R.~D. 2015, \apj, 808, 127, \dodoi{10.1088/0004-637X/808/2/127}

\bibitem[{{Gupta} \& {Bedell}(2024)}]{gupta24}
{Gupta}, A.~F., \& {Bedell}, M. 2024, \aj, 168, 29, \dodoi{10.3847/1538-3881/ad4ce6}

\bibitem[{{Hamann} {et~al.}(2019){Hamann}, {Montet}, {Fabrycky}, {Agol}, \& {Kruse}}]{hamann19}
{Hamann}, A., {Montet}, B.~T., {Fabrycky}, D.~C., {Agol}, E., \& {Kruse}, E. 2019, \aj, 158, 133, \dodoi{10.3847/1538-3881/ab32e3}

\bibitem[{Harris {et~al.}(2020)Harris, Millman, van~der Walt, Gommers, Virtanen, Cournapeau, Wieser, Taylor, Berg, Smith, Kern, Picus, Hoyer, van Kerkwijk, Brett, Haldane, del R{'{\i}}o, Wiebe, Peterson, G{'{e}}rard-Marchant, Sheppard, Reddy, Weckesser, Abbasi, Gohlke, \& Oliphant}]{numpy}
Harris, C.~R., Millman, K.~J., van~der Walt, S.~J., {et~al.} 2020, Nature, 585, 357, \dodoi{10.1038/s41586-020-2649-2}

\bibitem[{{Haywood} {et~al.}(2018){Haywood}, {Vanderburg}, {Mortier}, {Giles}, {L{\'o}pez-Morales}, {Lopez}, {Malavolta}, {Charbonneau}, {Collier Cameron}, {Coughlin}, {Dressing}, {Nava}, {Latham}, {Dumusque}, {Lovis}, {Molinari}, {Pepe}, {Sozzetti}, {Udry}, {Bouchy}, {Johnson}, {Mayor}, {Micela}, {Phillips}, {Piotto}, {Rice}, {Sasselov}, {S{\'e}gransan}, {Watson}, {Affer}, {Bonomo}, {Buchhave}, {Ciardi}, {Fiorenzano}, \& {Harutyunyan}}]{haywood18}
{Haywood}, R.~D., {Vanderburg}, A., {Mortier}, A., {et~al.} 2018, \aj, 155, 203, \dodoi{10.3847/1538-3881/aab8f3}

\bibitem[{{Hedges} {et~al.}(2020){Hedges}, {Angus}, {Barentsen}, {Saunders}, {Montet}, \& {Gully-Santiago}}]{hedges20}
{Hedges}, C., {Angus}, R., {Barentsen}, G., {et~al.} 2020, Research Notes of the American Astronomical Society, 4, 220, \dodoi{10.3847/2515-5172/abd106}

\bibitem[{Howard \& Fulton(2016)}]{howard16}
Howard, A.~W., \& Fulton, B.~J. 2016, Publications of the Astronomical Society of the Pacific, 128, 114401, \dodoi{10.1088/1538-3873/128/969/114401}

\bibitem[{Howard {et~al.}(2010)Howard, Johnson, Marcy, Fischer, Wright, Bernat, Henry, Peek, Isaacson, Apps, Endl, Cochran, Valenti, Anderson, \& Piskunov}]{howard10}
Howard, A.~W., Johnson, J.~A., Marcy, G.~W., {et~al.} 2010, The Astrophysical Journal, 721, 1467, \dodoi{10.1088/0004-637x/721/2/1467}

\bibitem[{Hunter(2007)}]{matplotlib}
Hunter, J.~D. 2007, Computing in Science \& Engineering, 9, 90, \dodoi{10.1109/MCSE.2007.55}

\bibitem[{{Isaacson} \& {Fischer}(2010)}]{isaacson10}
{Isaacson}, H., \& {Fischer}, D. 2010, \apj, 725, 875, \dodoi{10.1088/0004-637X/725/1/875}

\bibitem[{{Jeffreys}(1946)}]{jeffreys46}
{Jeffreys}, H. 1946, Proceedings of the Royal Society of London Series A, 186, 453, \dodoi{10.1098/rspa.1946.0056}

\bibitem[{{Kosiarek} \& {Crossfield}(2020)}]{kosiarek20}
{Kosiarek}, M.~R., \& {Crossfield}, I. J.~M. 2020, \aj, 159, 271, \dodoi{10.3847/1538-3881/ab8d3a}

\bibitem[{{Kosiarek} {et~al.}(2021){Kosiarek}, {Berardo}, {Crossfield}, {Laguna}, {Piaulet}, {Akana Murphy}, {Howell}, {Henry}, {Isaacson}, {Fulton}, {Weiss}, {Petigura}, {Behmard}, {Hirsch}, {Teske}, {Burt}, {Mills}, {Chontos}, {Mo{\v{c}}nik}, {Howard}, {Werner}, {Livingston}, {Krick}, {Beichman}, {Gorjian}, {Kreidberg}, {Morley}, {Christiansen}, {Morales}, {Scott}, {Crane}, {Wang}, {Shectman}, {Rosenthal}, {Grunblatt}, {Rubenzahl}, {Dalba}, {Giacalone}, {Villanueva}, {Liu}, {Dai}, {Hill}, {Rice}, {Kane}, \& {Mayo}}]{kosiarek21}
{Kosiarek}, M.~R., {Berardo}, D.~A., {Crossfield}, I. J.~M., {et~al.} 2021, \aj, 161, 47, \dodoi{10.3847/1538-3881/abca39}

\bibitem[{{Luque} \& {Pall{\'e}}(2022)}]{luque22}
{Luque}, R., \& {Pall{\'e}}, E. 2022, Science, 377, 1211, \dodoi{10.1126/science.abl7164}

\bibitem[{{Luque} {et~al.}(2019){Luque}, {Nowak}, {Pall{\'e}}, {Dai}, {Kaminski}, {Nagel}, {Hidalgo}, {Bauer}, {Lafarga}, {Livingston}, {Barrag{\'a}n}, {Hirano}, {Fridlund}, {Gandolfi}, {Justesen}, {Hjorth}, {Van Eylen}, {Winn}, {Esposito}, {Morales}, {Albrecht}, {Alonso}, {Amado}, {Beck}, {Caballero}, {Cabrera}, {Cochran}, {Csizmadia}, {Deeg}, {Eigm{\"u}ller}, {Endl}, {Erikson}, {Fukui}, {Grziwa}, {Guenther}, {Hatzes}, {Knudstrup}, {Korth}, {Lam}, {Lund}, {Mathur}, {Monta{\~n}es-Rodr{\'\i}guez}, {Narita}, {Nespral}, {Niraula}, {P{\"a}tzold}, {Persson}, {Prieto-Arranz}, {Quirrenbach}, {Rauer}, {Redfield}, {Reiners}, {Ribas}, \& {Smith}}]{luque19}
{Luque}, R., {Nowak}, G., {Pall{\'e}}, E., {et~al.} 2019, \aap, 623, A114, \dodoi{10.1051/0004-6361/201834952}

\bibitem[{{MacDougall} {et~al.}(2021){MacDougall}, {Petigura}, {Angelo}, {Lubin}, {Batalha}, {Beard}, {Behmard}, {Blunt}, {Brinkman}, {Chontos}, {Crossfield}, {Dai}, {Dalba}, {Dressing}, {Fulton}, {Giacalone}, {Hill}, {Howard}, {Huber}, {Isaacson}, {Kane}, {Mayo}, {Mo{\v{c}}nik}, {Akana Murphy}, {Polanski}, {Rice}, {Robertson}, {Rosenthal}, {Roy}, {Rubenzahl}, {Scarsdale}, {Turtelboom}, {Zandt}, {Weiss}, {Matthews}, {Jenkins}, {Latham}, {Ricker}, {Seager}, {Vanderspek}, {Winn}, {Brasseur}, {Doty}, {Fausnaugh}, {Guerrero}, {Henze}, {Lund}, \& {Shporer}}]{macdougall21}
{MacDougall}, M.~G., {Petigura}, E.~A., {Angelo}, I., {et~al.} 2021, \aj, 162, 265, \dodoi{10.3847/1538-3881/ac295e}

\bibitem[{{MacDougall} {et~al.}(2022){MacDougall}, {Petigura}, {Fetherolf}, {Beard}, {Lubin}, {Angelo}, {Batalha}, {Behmard}, {Blunt}, {Brinkman}, {Chontos}, {Crossfield}, {Dai}, {Dalba}, {Dressing}, {Fulton}, {Giacalone}, {Hill}, {Howard}, {Huber}, {Isaacson}, {Kane}, {Kosiarek}, {Mayo}, {Mo{\v{c}}nik}, {Akana Murphy}, {Pidhorodetska}, {Polanski}, {Rice}, {Robertson}, {Rosenthal}, {Roy}, {Rubenzahl}, {Scarsdale}, {Turtelboom}, {Tyler}, {Van Zandt}, {Weiss}, {Esparza-Borges}, {Fukui}, {Isogai}, {Kawauchi}, {Mori}, {Murgas}, {Narita}, {Nishiumi}, {Palle}, {Parviainen}, {Watanabe}, {Jenkins}, {Latham}, {Ricker}, {Seager}, {Vanderspek}, {Winn}, {Bieryla}, {Caldwell}, {Dragomir}, {Fausnaugh}, {Mireles}, \& {Rodriguez}}]{macdougall22}
{MacDougall}, M.~G., {Petigura}, E.~A., {Fetherolf}, T., {et~al.} 2022, \aj, 164, 97, \dodoi{10.3847/1538-3881/ac7ce1}

\bibitem[{{MacDougall} {et~al.}(2023){MacDougall}, {Petigura}, {Gilbert}, {Angelo}, {Batalha}, {Beard}, {Behmard}, {Blunt}, {Brinkman}, {Chontos}, {Crossfield}, {Dai}, {Dalba}, {Dressing}, {Fetherolf}, {Fulton}, {Giacalone}, {Hill}, {Holcomb}, {Howard}, {Huber}, {Isaacson}, {Kane}, {Kosiarek}, {Lubin}, {Mayo}, {Mo{\v{c}}nik}, {Akana Murphy}, {Pidhorodetska}, {Polanski}, {Rice}, {Robertson}, {Rosenthal}, {Roy}, {Rubenzahl}, {Scarsdale}, {Turtelboom}, {Tyler}, {Van Zandt}, {Weiss}, \& {Yee}}]{macdougall23}
{MacDougall}, M.~G., {Petigura}, E.~A., {Gilbert}, G.~J., {et~al.} 2023, \aj, 166, 33, \dodoi{10.3847/1538-3881/acd557}

\bibitem[{{NASA Exoplanet Archive}(2024)}]{nea}
{NASA Exoplanet Archive}. 2024, Planetary Systems, Version: 2024-09-02 17:46,  NExScI-Caltech/IPAC, \dodoi{10.26133/NEA12}

\bibitem[{{Otegi} {et~al.}(2020){Otegi}, {Bouchy}, \& {Helled}}]{otegi20a}
{Otegi}, J.~F., {Bouchy}, F., \& {Helled}, R. 2020, \aap, 634, A43, \dodoi{10.1051/0004-6361/201936482}

\bibitem[{{O'Toole} {et~al.}(2009){O'Toole}, {Tinney}, {Jones}, {Butler}, {Marcy}, {Carter}, \& {Bailey}}]{otoole09}
{O'Toole}, S.~J., {Tinney}, C.~G., {Jones}, H.~R.~A., {et~al.} 2009, \mnras, 392, 641, \dodoi{10.1111/j.1365-2966.2008.14051.x}

\bibitem[{pandas~development team(2020)}]{pandas}
pandas~development team, T. 2020, pandas-dev/pandas: Pandas, latest,  Zenodo, \dodoi{10.5281/zenodo.3509134}

\bibitem[{Polanski {et~al.}(2024)Polanski, Lubin, Beard, Murphy, Rubenzahl, Hill, Crossfield, Chontos, Robertson, Isaacson, Kane, Ciardi, Batalha, Dressing, Fulton, Howard, Huber, Petigura, Weiss, Angelo, Behmard, Blunt, Brinkman, Dai, Dalba, Fetherolf, Giacalone, Hirsch, Holcomb, Kosiarek, Mayo, MacDougall, MoÄnik, Pidhorodetska, Rice, Rosenthal, Scarsdale, Turtelboom, Tyler, Zandt, Yee, Coria, Dulz, Hartman, Householder, Lange, Langford, Louden, Siegel, Gilbert, Gonzales, Schlieder, Boyle, Christiansen, Clark, Fernandes, Lund, Savel, Gill, Beichman, Matson, Matthews, Furlan, Howell, Scott, Everett, Livingston, Ershova, Cheryasov, Safonov, Lillo-Box, Barrado, \& Morales-CalderÃ³n}]{polanski24}
Polanski, A.~S., Lubin, J., Beard, C., {et~al.} 2024, The Astrophysical Journal Supplement Series, 272, 32, \dodoi{10.3847/1538-4365/ad4484}

\bibitem[{{Rajpaul} {et~al.}(2017){Rajpaul}, {Buchhave}, \& {Aigrain}}]{rajpaul17}
{Rajpaul}, V., {Buchhave}, L.~A., \& {Aigrain}, S. 2017, \mnras, 471, L125, \dodoi{10.1093/mnrasl/slx116}

\bibitem[{{Rasmussen} \& {Williams}(2006)}]{rasmussen06}
{Rasmussen}, C.~E., \& {Williams}, C. K.~I. 2006, {Gaussian Processes for Machine Learning} (MIT Press)

\bibitem[{Ricker {et~al.}(2014)Ricker, Winn, Vanderspek, Latham, Bakos, Bean, Berta-Thompson, Brown, Buchhave, Butler, Butler, Chaplin, Charbonneau, Christensen-Dalsgaard, Clampin, Deming, Doty, Lee, Dressing, Dunham, Endl, Fressin, Ge, Henning, Holman, Howard, Ida, Jenkins, Jernigan, Johnson, Kaltenegger, Kawai, Kjeldsen, Laughlin, Levine, Lin, Lissauer, MacQueen, Marcy, McCullough, Morton, Narita, Paegert, Palle, Pepe, Pepper, Quirrenbach, Rinehart, Sasselov, Sato, Seager, Sozzetti, Stassun, Sullivan, Szentgyorgyi, Torres, Udry, \& Villasenor}]{ricker14}
Ricker, G.~R., Winn, J.~N., Vanderspek, R., {et~al.} 2014, Journal of Astronomical Telescopes, Instruments, and Systems, 1, 1 , \dodoi{10.1117/1.JATIS.1.1.014003}

\bibitem[{{Rodigas} \& {Hinz}(2009)}]{rodigas09}
{Rodigas}, T.~J., \& {Hinz}, P.~M. 2009, \apj, 702, 716, \dodoi{10.1088/0004-637X/702/1/716}

\bibitem[{{Rogers} {et~al.}(2023){Rogers}, {Schlichting}, \& {Owen}}]{rogers23}
{Rogers}, J.~G., {Schlichting}, H.~E., \& {Owen}, J.~E. 2023, \apjl, 947, L19, \dodoi{10.3847/2041-8213/acc86f}

\bibitem[{{Rogers} \& {Seager}(2010)}]{rogersSeager10}
{Rogers}, L.~A., \& {Seager}, S. 2010, \apj, 712, 974, \dodoi{10.1088/0004-637X/712/2/974}

\bibitem[{{Rosenthal} {et~al.}(2021){Rosenthal}, {Fulton}, {Hirsch}, {Isaacson}, {Howard}, {Dedrick}, {Sherstyuk}, {Blunt}, {Petigura}, {Knutson}, {Behmard}, {Chontos}, {Crepp}, {Crossfield}, {Dalba}, {Fischer}, {Henry}, {Kane}, {Kosiarek}, {Marcy}, {Rubenzahl}, {Weiss}, \& {Wright}}]{rosenthal21}
{Rosenthal}, L.~J., {Fulton}, B.~J., {Hirsch}, L.~A., {et~al.} 2021, \apjs, 255, 8, \dodoi{10.3847/1538-4365/abe23c}

\bibitem[{{Schwarz}(1978)}]{schwarz78}
{Schwarz}, G. 1978, Annals of Statistics, 6, 461

\bibitem[{{Shen} \& {Turner}(2008)}]{shen08}
{Shen}, Y., \& {Turner}, E.~L. 2008, \apj, 685, 553, \dodoi{10.1086/590548}

\bibitem[{{Tinetti} {et~al.}(2018){Tinetti}, {Drossart}, {Eccleston}, {Hartogh}, {Heske}, {Leconte}, {Micela}, {Ollivier}, {Pilbratt}, {Puig}, {Turrini}, {Vandenbussche}, {Wolkenberg}, {Beaulieu}, {Buchave}, {Ferus}, {Griffin}, {Guedel}, {Justtanont}, {Lagage}, {Machado}, {Malaguti}, {Min}, {N{\o}rgaard-Nielsen}, {Rataj}, {Ray}, {Ribas}, {Swain}, {Szabo}, {Werner}, {Barstow}, {Burleigh}, {Cho}, {Coud{\'e} du Foresto}, {Coustenis}, {Decin}, {Encrenaz}, {Galand}, {Gillon}, {Helled}, {Morales}, {Garc{\'\i}a Mu{\~n}oz}, {Moneti}, {Pagano}, {Pascale}, {Piccioni}, {Pinfield}, {Sarkar}, {Selsis}, {Tennyson}, {Triaud}, {Venot}, {Waldmann}, {Waltham}, {Wright}, {Amiaux}, {Augu{\`e}res}, {Berth{\'e}}, {Bezawada}, {Bishop}, {Bowles}, {Coffey}, {Colom{\'e}}, {Crook}, {Crouzet}, {Da Peppo}, {Sanz}, {Focardi}, {Frericks}, {Hunt}, {Kohley}, {Middleton}, {Morgante}, {Ottensamer}, {Pace}, {Pearson}, {Stamper}, {Symonds}, {Rengel}, {Renotte}, {Ade}, {Affer}, {Alard}, {Allard}, {Altieri}, {Andr{\'e}}, {Arena}, {Argyriou},
  {Aylward}, {Baccani}, {Bakos}, {Banaszkiewicz}, {Barlow}, {Batista}, {Bellucci}, {Benatti}, {Bernardi}, {B{\'e}zard}, {Blecka}, {Bolmont}, {Bonfond}, {Bonito}, {Bonomo}, {Brucato}, {Brun}, {Bryson}, {Bujwan}, {Casewell}, {Charnay}, {Pestellini}, {Chen}, {Ciaravella}, {Claudi}, {Cl{\'e}dassou}, {Damasso}, {Damiano}, {Danielski}, {Deroo}, {Di Giorgio}, {Dominik}, {Doublier}, {Doyle}, {Doyon}, {Drummond}, {Duong}, {Eales}, {Edwards}, {Farina}, {Flaccomio}, {Fletcher}, {Forget}, {Fossey}, {Fr{\"a}nz}, {Fujii}, {Garc{\'\i}a-Piquer}, {Gear}, {Geoffray}, {G{\'e}rard}, {Gesa}, {Gomez}, {Graczyk}, {Griffith}, {Grodent}, {Guarcello}, {Gustin}, {Hamano}, {Hargrave}, {Hello}, {Heng}, {Herrero}, {Hornstrup}, {Hubert}, {Ida}, {Ikoma}, {Iro}, {Irwin}, {Jarchow}, {Jaubert}, {Jones}, {Julien}, {Kameda}, {Kerschbaum}, {Kervella}, {Koskinen}, {Krijger}, {Krupp}, {Lafarga}, {Landini}, {Lellouch}, {Leto}, {Luntzer}, {Rank-L{\"u}ftinger}, {Maggio}, {Maldonado}, {Maillard}, {Mall}, {Marquette}, {Mathis}, {Maxted}, {Matsuo},
  {Medvedev}, {Miguel}, {Minier}, {Morello}, {Mura}, {Narita}, {Nascimbeni}, {Nguyen Tong}, {Noce}, {Oliva}, {Palle}, {Palmer}, {Pancrazzi}, {Papageorgiou}, {Parmentier}, {Perger}, {Petralia}, {Pezzuto}, {Pierrehumbert}, {Pillitteri}, {Piotto}, {Pisano}, {Prisinzano}, {Radioti}, {R{\'e}ess}, {Rezac}, {Rocchetto}, {Rosich}, {Sanna}, {Santerne}, {Savini}, {Scandariato}, {Sicardy}, {Sierra}, {Sindoni}, {Skup}, {Snellen}, {Sobiecki}, {Soret}, {Sozzetti}, {Stiepen}, {Strugarek}, {Taylor}, {Taylor}, {Terenzi}, {Tessenyi}, {Tsiaras}, {Tucker}, {Valencia}, {Vasisht}, {Vazan}, {Vilardell}, {Vinatier}, {Viti}, {Waters}, {Wawer}, {Wawrzaszek}, {Whitworth}, {Yung}, {Yurchenko}, {Zapatero Osorio}, {Zellem}, {Zingales}, \& {Zwart}}]{ariel}
{Tinetti}, G., {Drossart}, P., {Eccleston}, P., {et~al.} 2018, Experimental Astronomy, 46, 135, \dodoi{10.1007/s10686-018-9598-x}

\bibitem[{{Valencia} {et~al.}(2007){Valencia}, {Sasselov}, \& {O'Connell}}]{valencia07}
{Valencia}, D., {Sasselov}, D.~D., \& {O'Connell}, R.~J. 2007, \apj, 665, 1413, \dodoi{10.1086/519554}

\bibitem[{Van~Rossum \& Drake(2009)}]{python3}
Van~Rossum, G., \& Drake, F.~L. 2009, Python 3 Reference Manual (Scotts Valley, CA: CreateSpace)

\bibitem[{{Vogt} {et~al.}(1994){Vogt}, {Allen}, {Bigelow}, {Bresee}, {Brown}, {Cantrall}, {Conrad}, {Couture}, {Delaney}, {Epps}, {Hilyard}, {Hilyard}, {Horn}, {Jern}, {Kanto}, {Keane}, {Kibrick}, {Lewis}, {Osborne}, {Pardeilhan}, {Pfister}, {Ricketts}, {Robinson}, {Stover}, {Tucker}, {Ward}, \& {Wei}}]{vogt94}
{Vogt}, S.~S., {Allen}, S.~L., {Bigelow}, B.~C., {et~al.} 1994, in Society of Photo-Optical Instrumentation Engineers (SPIE) Conference Series, Vol. 2198, Instrumentation in Astronomy VIII, ed. D.~L. {Crawford} \& E.~R. {Craine}, 362, \dodoi{10.1117/12.176725}

\bibitem[{{Weiss} {et~al.}(2016){Weiss}, {Rogers}, {Isaacson}, {Agol}, {Marcy}, {Rowe}, {Kipping}, {Fulton}, {Lissauer}, {Howard}, \& {Fabrycky}}]{weiss16}
{Weiss}, L.~M., {Rogers}, L.~A., {Isaacson}, H.~T., {et~al.} 2016, \apj, 819, 83, \dodoi{10.3847/0004-637X/819/1/83}

\end{thebibliography}
